\documentclass[12pt,a4paper]{article}

\usepackage[usenames,dvipsnames]{xcolor}
\usepackage{jheppub}
\usepackage{url}
\usepackage{hyperref}
\usepackage{dsfont}
\usepackage{tensor}
\usepackage{braket}

\hypersetup{
    pdfencoding=unicode,
    colorlinks=true,
    urlcolor=Maroon,
    linkcolor=RoyalBlue,
    citecolor=Maroon,
    pdftitle={CFT characters from localization},
    pdfauthor={Lorenz Eberhardt, Duarte Fragoso and Hessel Posthuma},
    pdfdisplaydoctitle=true,
    pdfstartview=FitH,
    linktocpage=true
}

\newcommand{\be}{\begin{equation}}
\newcommand{\ee}{\end{equation}}
\newcommand\e{\mathrm{e}}
\renewcommand\d{\text{d}}
\DeclareMathOperator\tr{tr}
\allowdisplaybreaks[1]

\renewcommand\u{\mathfrak{u}}
\renewcommand\sl{\mathfrak{sl}}
\newcommand\su{\mathfrak{su}}

\newcommand\PSL{\text{PSL}}
\newcommand\PSU{\text{PSU}}
\newcommand\U{\text{U}}
\newcommand\SU{\text{SU}}
\newcommand\OSp{\text{OSp}}
\newcommand\Diff{\text{Diff}}
\newcommand\DiffS{\mathrm{Diff\,S}^1}
\newcommand\diag{\mathrm{diag}}
\newcommand\Ad{\mathrm{Ad}}
\newcommand\ad{\mathrm{ad}}
\newcommand\td{\mathrm{td}}

\newcommand\CC{\mathbb{C}}
\newcommand\ZZ{\mathbb{Z}}
\newcommand\RR{\mathbb{R}}
\newcommand\NN{\mathbb{N}}
\newcommand\CP{\mathbb{CP}}
\newcommand\DD{\mathbb{D}}

\newcommand\vac{\mathrm{vac}}
\newcommand\gen{\mathrm{gen}}
\newcommand\BPS{\mathrm{BPS}}
\newcommand\mR{\mathrm{R}}
\newcommand\out{\mathrm{out}}
\newcommand\bos{\mathrm{bos}}
\newcommand\ferm{\mathrm{ferm}}
\newcommand\red{\mathrm{red}}

\title{CFT characters from localization}
\author{Lorenz Eberhardt$^a$,}
\author{Duarte Fragoso$^{a,b}$ and}
\author{Hessel Posthuma$^c$}

\affiliation[a]{Institute for Theoretical Physics,
University of Amsterdam, Amsterdam, 1098XH, NL}
\affiliation[b]{Centro de Análise Matemática, Geometria e Sistemas Dinâmicos, \\
Departamento de Matemática, Instituto Superior Técnico, Universidade de Lisboa, \\
Av. Rovisco Pais 1, 1049-001 Lisbon, Portugal}
\affiliation[c]{Korteweg-de Vries Institute for Mathematics, University of Amsterdam, Amsterdam, 1098 XG, NL}
\emailAdd{l.eberhardt@uva.nl}
\emailAdd{duartefragoso@tecnico.ulisboa.pt}
\emailAdd{h.b.posthuma@uva.nl}

\abstract{We explain a general procedure to compute characters of chiral algebras of 2d (S)CFTs geometrically from their coadjoint orbits, in analogy with the coadjoint orbit method for finite-dimensional Lie groups. We demonstrate that this method can be used to give concise rederivations of the characters of the Virasoro algebra, Kac-Moody algebras, as well as supersymmetric generalizations.}

\begin{document}

\maketitle

\makeatletter
\g@addto@macro\bfseries{\boldmath}
\makeatother

\section{Introduction} \label{sec:introduction}
Characters are fundamental objects of representation theory. While being relatively simple objects, they often retain a useful amount of information about the underlying representation. 

A powerful method to study Lie group representations is via the orbit method \cite{Kirillov:2004rv}: Let $G$ act on the dual of its Lie algebra $\mathfrak{g}^*$ via the coadjoint action. The resulting coadjoint orbits $\mathcal{O}$ carry then both by construction a $G$-action, as well as an invariant symplectic form. They can thus be viewed as phase spaces for a classical system with a $G$-symmetry. Quantization of such a system gives a Hilbert space $\mathcal{H}_\mathcal{O}$ that, in turn, carries a $G$-representation. One can generate a large number of $G$-representations in this way and study their properties geometrically.

Quantization is not a very precise procedure, unless one specifies further data. A particularly simple case is when the phase space under consideration is a K\"ahler manifold, where one can apply K\"ahler quantization \cite{Woodhouse_Geometric_Quantization, Souriau:1997vk, Witten:1987ty}, which formalizes the coherent state quantization of quantum mechanics \cite{Rawnsley:1976gb, Perelomov:1971bd, Nair:2018use}. Essentially, wavefunctions become holomorphic sections of a line bundle $\mathcal{L}$ such that the curvature is the symplectic form $\omega$, i.e.\ $c_1(\mathcal{L})=[\omega]$ in cohomological terms. In this case, we thus have $\mathcal{H}_\mathcal{O}=H^0(\mathcal{O},\mathcal{L})$.

In particular, coadjoint orbits happen to be K\"ahler for compact semisimple Lie groups, and the description of the irreducible representations as holomorphic sections is the content of the Borel--Weil--Bott theorem \cite{Serre:1954bw, Bott:1957hb, BWBTits}. For example, in the simple case of $G=\SU(2)$, the Lie algebra is three-dimensional and coadjoint orbits are two-spheres since only the distance to the origin is preserved under the $\SU(2)$ action. The $n+1$-dimensional irreducible representation of $\SU(2)$ can then be realized on the sections $H^0(\CP^1,\mathcal{O}(n))$, which can be identified with homogeneous polynomials of degree $n$ in two variables \cite{Kirillov:2004rv}.

For the purpose of computing characters, one is interested in computing the trace of a group element over $H^0(\mathcal{O},\mathcal{L})$. A convenient tool for this computation is given by the equivariant version of the Grothendieck-Riemann-Roch index theorem \cite{vergne2016equivariantriemannrochtheoremgraded, Berline:2004hd}, which under the assumptions of certain vanishing theorems gives
\be 
\chi_\mathcal{O}(\mathrm{e}^X)=\tr_{H^0(\mathcal{O},\mathcal{L})} (\mathrm{e}^{X})=\int_{\mathcal{O}} \td^G(\mathcal{O})\, \mathrm{e}^{c_1^G(\mathcal{L})}(X)\ ,
\ee
with $X \in \mathfrak{g}$ and where $\td^G$ and $c_1^G$ denote equivariant completions of the Todd and Chern classes, respectively. The final tool one needs is equivariant localization \cite{ Atiyah_Bott, Witten:1992xu, Pestun:2016zxk}, which reduces the integral over the orbit to a sum over the fixed points of the action. The upshot is that the character can actually be computed by the knowledge of the fixed points of the $G$-action on the coadjoint orbit, as well as the local action. Therefore, characters can be computed completely geometrically.
\bigskip

This beautiful picture is well-studied in the context of finite-dimensional Lie groups, where it can be used to give an elegant derivation of Weyl's character formula \cite{Pestun:2016zxk}. In this paper, we shall be primarily interested in characters appearing in two-dimensional conformal field theory, such as characters of the Virasoro algebra, Kac-Moody algebras relevant to WZW models, as well as supersymmetric extensions \cite{Kac:1990gs, loopgroup_Pressley_Segal, DiFrancesco:1997nk, Kiritsis:1986rv, Eguchi:1988af}. In this case, the orbits often still turn out to carry a formal K\"ahler structure, even though they are now infinite-dimensional objects. Even though this makes the construction lose its mathematical rigor, the logical chain still applies and one can compute the relevant characters from equivariant localization \cite{Alekseev:2020jja, Murthy:2025ioh}. The final results for the characters can be found in eqs.~\eqref{eq:generic-Virasoro-character}, \eqref{eq:vacuum-Virasoro-character}, \eqref{eq:kac-weyl-general}, \eqref{eq:N1-Virasoro-characters}, \eqref{eq:N2-Virasoro-generic}, \eqref{eq:N2-Virasoro-special}, \eqref{eq:n4-generic-character} and \eqref{eq:n4-bps-character}.

This method gives one a rather nice and uniform way to compute various CFT characters that in the literature were often derived in a somewhat laborious way by studying the precise null-vector structure of the relevant modules. We will also explain that the method can be extended to characters of algebras involving fermions, where the relevant coadjoint orbits become supermanifolds \cite{Cotler:2018zff, Yang:1991rd}. We exemplify the method by computing the characters of the Virasoro algebra, the Kac-Moody algebra, as well as the Virasoro algebra with $\mathcal{N}=1$, $2$, and small $\mathcal{N}=4$ supersymmetry.

This procedure is also related to three-dimensional theories, such as Chern-Simons theory and (super) gravity in three dimensions. Quantization of Chern-Simons theory on the punctured disk is precisely equivalent to the quantization of a coadjoint orbit of the Kac-Moody algebra, with similar statements holding for gravity \cite{Witten:1988hf, loopgroup_Pressley_Segal, Witten:1987ty, Eberhardt:2022wlc, Alekseev_Shatashvilli_PI}. Therefore, the Hilbert space that one analyzes here is by construction the Hilbert space of the punctured disk and the character is the partition function of a solid torus with an embedded Wilson line around the non-contractible cycle. As such, the localization procedure in the Kac-Moody case is a special case of the localization techniques employed in Chern-Simons theory \cite{Beasley:2009mb, Beasley:2005vf, Marino:2011nm}.
\bigskip

This paper is organized as follows. We start in Section~\ref{sec:finite-dimensional-case} by explaining the coadjoint orbit method, as well as the index theorem and equivariant localization. We then apply it to a finite-dimensional compact simple Lie group and demonstrate that it reproduces the Weyl character formula. In Section~\ref{sec:application-to-cft}, we apply the method to 2d CFT, first in the case of the Virasoro algebra, then for the Kac-Moody algebra and finally for the $\mathcal{N}=1$, $\mathcal{N}=2$, and $\mathcal{N}=4$ super Virasoro cases. We include two technical appendices \ref{app:reducing-superindex-theorem} and \ref{app:canonical-line-bundles}.

\section{The finite-dimensional case} \label{sec:finite-dimensional-case}
In this Section, we start by reviewing geometric quantization of coadjoint orbits of finite-dimensional Lie groups. Equivariant localization applied to coadjoint orbits gives an elegant way to rederive Weyl's character formula. An analogous derivation is given in \cite{Atiyah:1968ms,Alekseev:2020jja}.
 
\subsection{Coadjoint orbits and geometric quantization} \label{subsec:coadjoint-orbits-geometric-quantization}
A powerful approach to the representation theory of Lie groups, due to Kirillov, Kostant, and Souriau, is the orbit method \cite{Kirillov:2004rv,Kostant:1970vx,Souriau:1997vk}. It proceeds as follows. First, the coadjoint orbits $\mathcal{O}_\lambda$ of a compact, semisimple Lie group $G$ carry K\"ahler structures. Quantization assigns a Hilbert space to such a coadjoint orbit. For a K\"ahler manifold, this Hilbert space can be identified with a space of holomorphic sections of a line bundle $\mathcal{L}_\lambda$ on $\mathcal{O}_\lambda$, i.e.\ $H^0(\mathcal{O}_\lambda, \mathcal{L}_\lambda)$. The content of the Borel--Weil--Bott theorem \cite{Serre:1954bw,Bott:1957hb} is that, for a dominant integral weight $\lambda$, $H^0(\mathcal{O}_\lambda, \mathcal{L}_\lambda)$ defines precisely an irreducible representation of the group $G$ and, for $G$ compact and semisimple, all finite-dimensional irreducible representations arise in this way. We start by giving more details on coadjoint orbits.

\paragraph{Coadjoint orbits.} Let $G$ be a Lie group. The adjoint action $\Ad$ naturally defines a representation on its Lie algebra. The coadjoint action on the Lie algebra dual $\mathfrak{g}^*$ is defined as
\be
 \langle \mathrm{Ad}^*(g)\lambda, X \rangle :=\langle \lambda, \mathrm{Ad}(g^{-1})X\rangle\ ,
\ee
where $g \in G$, $\lambda \in \mathfrak{g}^*$, and the bracket denotes the usual pairing between the Lie algebra and its dual. Coadjoint orbits $\mathcal{O}_\lambda$ are then the orbits of this action through some element $\lambda \in \mathfrak{g}^*$, which by the orbit--stabilizer theorem are isomorphic to homogeneous spaces of the form $G/H$, where $H$ is the stabilizer group of $\lambda$ with respect to the coadjoint action.\par

Coadjoint orbits naturally carry a symplectic structure, which is given by the \textit{Kirillov--Kostant--Souriau} (\textit{KKS}) form \cite{Kirillov:2004rv,Kostant:1970vx,Souriau:1997vk}. Let us describe it at the point $\lambda$. Tangent vectors to $\mathcal{O}_\lambda$ at $\lambda$ are obtained by infinitesimal variations of $\lambda$ of the form $\xi_X(\lambda):=\ad^*(X)\lambda$ for $X \in \mathfrak{g}$, and where $\ad^*$ is the Lie algebra action corresponding to $\Ad^*$. We thus set
\be
 \omega_{\lambda}(\xi_X(\lambda),\xi_Y(\lambda)) := \bigl\langle \lambda,\,[X,Y]\bigr\rangle\ . \label{eq:kks-symplectic-form}
\ee
One can check that this is well-defined, i.e.\ independent of the choice of $X$ and $Y$. It can also be transported to any other point in $\mathcal{O}_\lambda$ via the $G$-action. $\omega$ defines a nondegenerate closed two-form, which turns $\mathcal{O}_\lambda$ into a symplectic manifold.

The $G$-action is Hamiltonian: there is a $G$-equivariant momentum map $\mu: \mathcal{O}_\lambda \to \mathfrak{g}^*$ such that for $X \in \mathfrak{g}$
\be 
\omega(\xi_X,\bullet)=\d(\langle \mu, X \rangle )\ , \label{eq:momentum-map} 
\ee
where $\xi_X$ is the vector field on $\mathcal{O}_\lambda$ realizing the infinitesimal action of $X \in \mathfrak{g}$. Thus, $\mu$ contains all the information about the $G$-action. For coadjoint orbits, $\mu:\mathcal{O}_\lambda \hookrightarrow \mathfrak{g}^*$ is just the inclusion.

\paragraph{Generalities about quantization.} $\mathcal{O}_\lambda$ is a symplectic manifold with a $G$-action. It therefore describes a classical mechanics system, up to the choice of a Hamiltonian function. Irrespective of this choice, one can try to apply quantization to produce a quantum mechanical system with a $G$-symmetry, i.e.\ in particular a Hilbert space $\mathcal{H}_\lambda$ that carries a $G$-representation. Quantization of a general symplectic manifold is not straightforward. The general obstruction is that one has to choose a polarization, meaning a choice of half of the coordinates on which the wavefunction will depend. In general, no natural choice exists. However, it turns out that many of the coadjoint orbits are complex and even K\"ahler manifolds. In that case, a natural choice is to let the wavefunction depend holomorphically on the coordinates. This goes under the name of K\"ahler quantization.

In K\"ahler quantization, one first has to pick a holomorphic line bundle with a hermitian metric (the prequantum line bundle) $\mathcal{L}_\lambda$ with $c_1(\mathcal{L}_\lambda)=[\omega]$ (with $\omega$ normalized so that $[\omega]$ rather than $\bigl[\tfrac{\omega}{2\pi}\bigr]$ is integral; this avoids stray factors of $2\pi$ in the character formulas). This requires that $[\omega]$ takes values in integer cohomology, which is the quantization condition \cite{Woodhouse_Geometric_Quantization}. The choice of such line bundle is unique in our case since coadjoint orbits of semisimple compact Lie groups are simply connected. Wavefunctions in the quantum theory become polarized, i.e.\ holomorphic sections of this line bundle (with respect to the Chern connection). Thus, the Hilbert space is identified with $H^0(\mathcal{O}_\lambda,\mathcal{L}_\lambda)$, which then carries a natural $G$-action.

\paragraph{Criterion for K\"ahler structures.} To proceed, it is necessary to analyze which coadjoint orbits can be endowed with a K\"{a}hler structure. A criterion for the existence of such structures goes back to \cite{Borel_Kahler} in the compact case and was extended to the Virasoro setting in \cite{Witten:1987ty}, which we now recall.
Consider a coadjoint orbit $\mathcal{O}_\lambda\cong G/H$ and let $\mathfrak{h}$ be the Lie algebra of the stabilizer $H$. Then, the tangent space of the orbit at the identity coset can be identified with a vector-space complement $\mathcal{K}$ of $\mathfrak{h}$ in $\mathfrak{g}$, which we choose $H$-invariant (e.g.\ the Killing-orthogonal complement for $G$ compact),
\be
\mathfrak{g}=\mathfrak{h} \oplus \mathcal{K}\ .
\ee
Providing a complex structure to the coadjoint orbit is done by choosing a splitting of the complexification of the complement,
\be
\mathcal{K}_\CC=\mathcal{K}^+ \oplus \mathcal{K}^-\ ,
\ee
such that $\mathcal{K}^-$ is the complex conjugate of $\mathcal{K}^+$.

Beyond that, the conditions for the splitting to be $H$-invariant (so that the corresponding complex structure on $G/H$ is $G$-invariant) and integrable are, respectively,
\begin{subequations}
\begin{align}
&[\mathfrak{h}, \mathcal{K}^+]\subset  \mathfrak{h} \oplus \mathcal{K}^+\ , \label{eq:hkp-criterion}\\
&[\mathcal{K}^+, \mathcal{K}^+] \subset \mathfrak{h} \oplus \mathcal{K}^+ \ ,\label{eq:kpkp-criterion}
\end{align} \label{eq:witten-criterion}%
\end{subequations}
and similarly for $\mathcal{K}^-$.
This criterion will be used extensively in the following sections.

For regular coadjoint orbits, i.e. where $H=T$, a maximal torus, the choice of splitting $\mathcal{K}_\CC=\mathcal{K}^+ \oplus \mathcal{K}^-$ amounts to a choice of positive roots. Such a choice is not unique, but only for one the resulting invariant complex structure satisfies the K\"ahler condition.

With the tools detailed above, we now have all the ingredients to apply Kirillov's orbit method to systematically construct unitary representations of Lie groups. We first identify all quantizable coadjoint orbits and use the criterion \eqref{eq:witten-criterion} to single out those admitting a K\"{a}hler structure. Then, we use the K\"{a}hler polarization to find the quantum Hilbert space of holomorphic sections of the prequantum line bundle, $\mathcal{H}_{\lambda}=H^0(\mathcal{O}_\lambda,\mathcal{L}_\lambda)$. Since this line bundle inherits the group action from the underlying orbit, the quantum Hilbert space is a representation of the original Lie group.
 
\subsection{Index theorem} \label{subsec:index-theorem}
In the following, we shall be mostly interested in the character of the representation, not the actual representation itself. The character can be computed directly from the orbit.

\paragraph{Euler characteristic and dimension.} 
Let us begin by computing the dimension of 
$\mathcal{H}_\lambda=H^0(\mathcal{O}_\lambda,\mathcal{L}_\lambda)$. This is, in general, a difficult problem. However, the Hirzebruch--Riemann--Roch theorem computes a closely related quantity from the geometry of $\mathcal{O}_\lambda$ \cite{Hirzebruch:1966sn}. It computes the alternating sum of the dimensions of higher cohomology groups obtained from the Dolbeault complex
\be
\bar{\partial}:  \Omega^{0,p}(\mathcal{O}_\lambda,\mathcal{L}_\lambda) \to \Omega^{0,p+1}(\mathcal{O}_\lambda,\mathcal{L}_\lambda)\ .
\ee
It states
\be
\sum_{m=0}^{\dim_{\CC} \mathcal{O}_\lambda}(-1)^m \dim H^m(\mathcal{O}_\lambda, \mathcal{L}_\lambda)=\int_{\mathcal{O}_\lambda}\td(T\mathcal{O}_\lambda)\,\e^{c_1(\mathcal{L}_\lambda)}\ , \label{eq:hrr-theorem}
\ee
with $T \mathcal{O}_\lambda$ the holomorphic tangent bundle.
In K\"ahler quantization, the higher cohomology groups vanish, $H^m(\mathcal{O}_\lambda,\mathcal{L}_\lambda)=0$ for $m>0$. This follows from the Kodaira vanishing theorem, which asserts that on a compact K\"ahler manifold $X$, $H^q(X,\mathcal{K}_X\otimes\mathcal{M})=0$ for $q>0$ whenever $\mathcal{M}$ is an ample line bundle; here $\mathcal{K}_X=\det T^*X$ denotes the canonical bundle. To apply this to $\mathcal{L}_\lambda$, write $\mathcal{L}_\lambda=\mathcal{K}_{\mathcal{O}_\lambda}\otimes(\mathcal{K}_{\mathcal{O}_\lambda}^{-1}\otimes\mathcal{L}_\lambda)$ and remark that for coadjoint orbits of compact semisimple Lie groups the anticanonical bundle $\mathcal{K}_{\mathcal{O}_\lambda}^{-1}$ is already ample, see e.g.\ \cite[Prop 1.4.1 and Prop 2.2.7]{Brion:2004fla}. Combined with the (semi-)positivity of $c_1(\mathcal{L}_\lambda)=[\omega]$ (represented by the K\"ahler form), this implies that $\mathcal{K}_{\mathcal{O}_\lambda}^{-1}\otimes\mathcal{L}_\lambda$ is ample. Kodaira vanishing then yields the claim. 

Thus, the alternating sum on the left-hand side reduces to the dimension of the quantum Hilbert space $\mathcal{H}_\lambda =H^0(\mathcal{O}_\lambda,\mathcal{L}_\lambda)$. Moreover, the right-hand side simplifies since, by construction, $c_1(\mathcal{L}_\lambda)=[\omega]$, giving
\be
\dim \mathcal{H}_\lambda=\int_{\mathcal{O}_\lambda}\td(T\mathcal{O}_\lambda)\,\e^{\omega}\ . \label{eq:dimension-index-theorem}
\ee
\paragraph{Character and equivariant cohomology.} The character of a representation $\rho$ is the function $\chi(g)=\tr\rho(g)$; evaluated at $g=\e^X$ with $X \in \mathfrak{g}$, it specializes to the dimension at $X=0$. One can generalize \eqref{eq:dimension-index-theorem} to account for the $G$-action on $\mathcal{O}_\lambda$ by considering the equivariant version of HRR \cite{Berline:2004hd}. Substituting the integrated characteristic classes with their equivariantly closed extensions, the dimension of the Hilbert space is promoted to a character,
\be 
\chi_{\lambda}(\e^X)=\int_{\mathcal{O}_\lambda} \big(\td^G (T\mathcal{O}_\lambda)\,\e^{\omega^G}\big)(X)\ ,\quad \text{for }X \in \mathfrak{g}\ . \label{eq:character-equivariant-index-theorem}
\ee
Here, the superscript $G$ denotes that these classes are elements of equivariant cohomology. 
Let us recall the definition of equivariant cohomology in terms of the Cartan model. Equivariant differential forms are $G$-equivariant polynomial maps $\alpha:\mathfrak{g}\to\Omega^*(M)$, i.e.\ elements of $(\Omega^*(M)\otimes S^*\mathfrak{g}^*)^G$, with $X\in\mathfrak{g}$ assigned degree $2$. They are endowed with the equivariant differential
\be
(\d^G\alpha)(X)=\d (\alpha(X))-\iota_{\xi_X}\alpha(X)\ , \label{eq:equivariant-differential}
\ee
where $\xi_X$ is the vector field generated by $X$. The equivariant extension of the KKS symplectic form is given, for $X \in \mathfrak{g}$, by
\be
\omega^G(X)=\omega+\mu(X)\ , \label{eq:equivariant-symplectic-form} 
\ee
which is closed with respect to the equivariant differential $\d^G$ thanks to \eqref{eq:momentum-map}.

The advantages of using equivariant cohomological forms are twofold. Not only do we obtain the character of the representation as a generalization of its dimension, but we also express it as an integral which can be localized since the coadjoint orbit carries a natural $G$-action.

\subsection{Localization} \label{subsec:localization}
At first sight, the equivariant formula \eqref{eq:character-equivariant-index-theorem} looks more complicated than the bare index theorem \eqref{eq:dimension-index-theorem}. However, it is actually simpler to evaluate thanks to equivariant localization \cite{Duistermaat_Heckman, Atiyah_Bott}. 

\paragraph{Equivariant localization.}
At its core, localization allows one to rewrite an integral of an equivariantly closed top form on a manifold $M$ carrying a $G$-action as an integral over $M^T$, the fixed-point set of a maximal torus $T \subset G$. For physicists, this method is well-known from supersymmetric localization in gauge theories \cite{Pestun:2007rj,Pestun:2016zxk}. Let us briefly sketch the argument for a manifold $M$ with a $\U(1)$ action.

Consider an equivariant closed form $\alpha$ which we want to integrate over $M$,
\be
I[\alpha]=\int_M \alpha\ .
\ee
Here, the right hand side can be viewed as a polynomial function on $\mathfrak{g}$, i.e. an element in $S\mathfrak{g}^*$.
Considering $\beta$, a $\U(1)$-invariant $1$-form on $M$, one can deform the original integral,
\be
I_t[\alpha]=\int_M \alpha\, \e^{t\,\d^{\U(1)}\beta}\ ,
\ee
where $t \in \RR$ is the deformation parameter. Since $\d^{\U(1)}\beta$ is equivariantly exact (BRST exact in the supersymmetric gauge theory setting), the integral does not depend on $t$. Localization follows from considering the $t \to \infty$ limit of the deformed integral, which suppresses the integral away from the locus where $\d^{\U(1)}\beta=0$. A useful appropriate choice for $\beta$ is
\be
\beta=g(\xi,\,\bullet\,)\ ,
\ee
with $\xi$ the fundamental vector field generating the $\U(1)$ action and $g$ a $\U(1)$-invariant metric. With this choice, 
\be 
\d^{\U(1)}\beta=\d \beta-\iota_\xi \beta=\d \beta-g(\xi,\xi)\ ,
\ee 
so the vanishing locus of the degree zero piece of $\d^{\U(1)}\beta$ is precisely the fixed-point set of the action. To get the full localization formula, one also has to evaluate the integral over the one-loop fluctuations around the fixed point. In physics lingo, it is one-loop exact because higher loops are suppressed by $t$. This gives a factor of $\det^{-1/2}$ from the Gaussian integral. In cohomology, this one-loop determinant is described by the equivariant Euler class $e^{\U(1)}(\mathcal{N})$ of the normal bundle $\mathcal{N}$ to the fixed-point submanifold $M^{\U(1)}$ within $M$. Concretely, via Chern--Weil theory, the equivariant Euler class is represented by the Pfaffian of the equivariant curvature 2-form. 

For more general group actions $G$, we localize with respect to a maximal torus $T \subset G$. This is the content of the Atiyah--Bott--Berline--Vergne (ABBV) formula \cite{Atiyah_Bott, Berline:2004hd},
\be
\int_M \alpha = \int_{M^T} \frac{\alpha}{e^T(\mathcal{N})}\ . \label{eq:abbv}
\ee
\paragraph{Application to the index theorem.} Let us return to our original goal of computing the integral in \eqref{eq:character-equivariant-index-theorem}. The coadjoint orbit carries a natural $G$-action, which restricts to an action of the maximal torus $T \subset G$. A direct application of the abelian ABBV formula then equates the character to an integral over the fixed-point set of equivariant characteristic classes. As we shall see below, the $T$-fixed-point set of $\mathcal{O}_\lambda$ consists of a number of isolated points. 
Thus, the integral reduces to a sum over each fixed point of the degree-zero contributions of all the involved equivariant characteristic classes: the contribution of the equivariant symplectic form reduces to the momentum map $\mu(p)$, see \eqref{eq:equivariant-symplectic-form}. Moreover, since the normal bundle to a point is simply the tangent space at that point, the character can be rewritten as a fixed-point formula
\be \label{eq:localized-character}
\chi_{\lambda}=\sum_{p \in \mathcal{O}_\lambda^T} \frac{\td_p (T\mathcal{O}_\lambda)\,\e^{\mu(p)}}{e_p(T\mathcal{O}_\lambda)}\ ,
\ee
where $\e^{\mu(p)}$ is shorthand for $X\mapsto \e^{\langle\mu(p),X\rangle}$, and the equivariant Todd and Euler classes at $p$ are functions of $X\in\mathfrak{g}$.
 
It remains to write expressions for the Todd and Euler classes. For a line bundle $\mathcal{L}$, the only independent equivariant characteristic class is its equivariant first Chern class $x=c_1(\mathcal{L})$. The Todd class and Euler class are defined respectively as
\be 
\td_p(\mathcal{L})=\frac{x}{1-\e^{-x}}\ , \qquad e_p(\mathcal{L})=x\ ,
\ee
so the linear term in~\eqref{eq:localized-character} cancels out. We can decompose the holomorphic tangent space $T_p \mathcal{O}_\lambda$ into eigenspaces of $T \subset G$, $T_p\mathcal{O}_\lambda \cong \mathcal{L}_1 \oplus \cdots \oplus \mathcal{L}_n$. Since the classes behave multiplicatively, we can write
\begin{equation}
\frac{\td_p(T\mathcal{O}_\lambda)}{e_p(T \mathcal{O}_\lambda)}=\prod_{i=1}^n  \frac{1}{1-\e^{-x_i}}\ , \label{eq:equivariant-todd-euler-ratio}
\end{equation}
where $x_i=c_1(\mathcal{L}_i)$ are the equivariant Chern roots at $p$, i.e.\ the eigenvalues of the infinitesimal $T$-action on the holomorphic directions of the complexified tangent space of $\mathcal{O}_\lambda$ at $p$.

\subsection{Weyl character formula} \label{subsec:weyl-character-formula}
We will now specialize to compact semisimple simply connected finite-dimensional Lie groups. In this case, the localized character \eqref{eq:localized-character} reproduces the Weyl character formula upon further analysis. 
We will first consider the generic orbits of type $G/T$.

\paragraph{Quantization condition.} The quantization condition requires $[\omega]$ to take values in integer cohomology $H^2(G/T,\ZZ)$. For $G=\SU(2)$ and $T=\U(1)$, $G/T \cong \CP^1$. Line bundles on $\CP^1$ are fully classified by the first Chern class. Suppose $[\omega]$ is the $n$-th multiple of the generator of $H^2(\CP^1,\ZZ)$. We can identify $n=\langle \lambda,\alpha^\vee \rangle \in \ZZ$, where $\alpha^\vee \in \su(2)$ is the simple coroot associated to the simple root $\alpha$. Recall that the coroot $\alpha^\vee \in \mathfrak{t}$ can be defined as $\alpha^\vee=[E_\alpha,F_\alpha]$, where $E_\alpha \in \mathfrak{g}_\alpha$ and $F_\alpha \in \mathfrak{g}_{-\alpha}$ are normalized so that $\langle\alpha,\alpha^\vee\rangle = 2$. The normalization is chosen such that integer values of $\langle\lambda,\alpha^\vee\rangle$ correspond exactly to weights, as can be checked e.g.\ for the fundamental representation with weight $\lambda = \varpi$, where $\langle\varpi,\alpha^\vee\rangle = 1$.

The corresponding line bundle is $\mathcal{L}_\lambda\cong\mathcal{O}(n)$, which is well-defined iff $n=\langle\lambda,\alpha^\vee\rangle\in\ZZ$ -- this is the quantization condition. For $n\geq 0$, the space of holomorphic sections is $(n+1)$-dimensional, spanned by homogeneous polynomials of degree $n$ in two variables, and yields the corresponding irreducible representation of $\SU(2)$.

For the general case, we can consider the $\SU(2)$ subgroup of $G$ associated to a simple root $\alpha$, obtained by exponentiating the $\sl_2$ triple $E_\alpha$, $F_\alpha$, and $\alpha^\vee$ associated to the simple root. Let $\iota_\alpha:\SU(2) \longrightarrow G$ be the embedding. Then $\mathrm{S}^2_\alpha=\iota_\alpha(\SU(2)) \cdot T \subset G/T$ is an embedded two-sphere. In fact, $\{[\mathrm{S}^2_\alpha]\}_{\alpha \text{ simple}}$ is a $\ZZ$-basis for $H_2(G/T,\ZZ)$. This is part of the Bruhat/Schubert cell decomposition, see e.g.\ \cite{Brion:2004fla}. 

We will now show that the pullback of the KKS symplectic form on $G/T$ gives the KKS symplectic form on $\SU(2)/\U(1)$. From the definition \eqref{eq:kks-symplectic-form}, we have at the basepoint $eT \in G/T$,
\begin{multline}
(\iota_\alpha^*\omega_\lambda)(x,y)=\omega_\lambda(\d\iota_\alpha(x),\d\iota_\alpha(y))=\big \langle \lambda, [\d\iota_\alpha x,\d\iota_\alpha y] \big \rangle\\
=\big \langle \lambda,\d\iota_\alpha [x,y] \big \rangle=\big \langle (\d\iota_\alpha)^*\lambda, [x,y] \big \rangle=\omega_{(\d\iota_\alpha)^*\lambda}(x,y)\ , \label{eq:kks-reduction}
\end{multline}
where $x,y \in \su(2)$. Here $\d \iota_\alpha:\su(2) \longrightarrow \mathfrak{g}$ is the induced map of the inclusion $\iota_\alpha$ on the Lie algebra and $(\d\iota_\alpha)^*:\mathfrak{g}^* \longrightarrow \su(2)^*$ its dual. Equality at other points follows because both forms are $\SU(2)$-invariant, and an invariant 2-form on a homogeneous space is determined by its value at any single point.

Thus, $(\d\iota_\alpha)^*\lambda$ needs to be integral for every choice of $\alpha$. In view of the above discussion, this means that $\langle (\d\iota_\alpha)^*\lambda,\alpha_{\su(2)}^\vee \rangle=\langle \lambda, \d\iota_\alpha (\alpha^\vee_{\su(2)}) \rangle=\langle \lambda,\alpha^\vee\rangle \in \ZZ$. Here $\alpha^\vee_{\su(2)} \in \mathfrak{t}_{\su(2)}$ is the simple coroot in the $\su(2)$ algebra. It is mapped to the coroot $\alpha^\vee\in \mathfrak{t}$ by the embedding $\d \iota_\alpha$. Thus the quantization condition gives $\langle \lambda,\alpha^\vee \rangle \in \ZZ$, which is precisely the requirement that $\lambda$ is in the weight lattice $P$ of $\mathfrak{g}$.

\paragraph{Fixed-point set.} Let us describe the fixed-point set of the $T$-action on orbits of the type $G/T$. Let $gT\in G/T$ be an element of the coadjoint orbit (identified with $\mathcal{O}_\lambda$ via $\Ad^*(gT) \lambda=\Ad^*(g)\lambda$ since $T$ is the stabilizer of $\lambda$) fixed by $t \in T$. Then
\be
tgT = gT \iff g^{-1}tg \in T\ .
\ee
Since we want $gT$ to be fixed by all $t \in T$, we obtain the fixed-point set
\be
W \cong \{g \in G \mid g^{-1} t g \in T\text{ for all }t \in T\}/T\ .
\ee
Note that $g^{-1}Tg$ is itself a maximal torus contained in $T$; since both have the same dimension and are connected, $g^{-1}Tg=T$, i.e.\ $g\in N_G(T)$. The fixed-point set is therefore $N_G(T)/T$, which is, by definition, the Weyl group $W$. We conclude that every fixed point is of the type $wT$ for some $w \in W$, and since the momentum map is the inclusion $\mathcal{O}_\lambda \hookrightarrow \mathfrak{g}^*$, it evaluates there to $\mu(w)=w(\lambda)$.
For example, in the case of $G=\SU(N)$, $T$ can be taken to be the diagonal matrices in $\SU(N)$ and $W \cong S_N$ consists of permutation matrices.

\paragraph{The character formula.} To conclude the computation of the character formula \eqref{eq:localized-character}, one needs to evaluate the equivariant Chern roots of the tangent bundle of the coadjoint orbit. Consider the usual root space decomposition of the Lie algebra of $G$,
\be
\mathfrak{g}_\CC = \mathfrak{t}_\CC \oplus \bigoplus_{\alpha \in \Delta} \mathfrak{g}_{\alpha}\ ,
\ee
where $\mathfrak{t}$ is the Lie algebra of $T$, $\Delta$ is the root system of $\mathfrak{g}$, and $\mathfrak{g}_\alpha$ is the root space of $\alpha$. The tangent space of the orbit at the identity coset $eT \in G/T$ can then be identified with $\bigoplus_{\alpha \in \Delta}\mathfrak{g}_\alpha$. Choosing a complex structure on $G/T$ is therefore equivalent to choosing a decomposition of $\Delta$ into positive roots $\Delta^+$ and negative roots $\Delta^-=\Delta \setminus \Delta^+$, with $\bigoplus_{\alpha \in \Delta^+}\mathfrak{g}_\alpha$ chosen as the holomorphic tangent space at the identity coset. The equivariant Chern roots at the identity coset are thus the positive roots, and at any other fixed point $w\in W$ they are given by applying $w$ to $\Delta^+$.\par
Substituting into \eqref{eq:localized-character} with $\mu(w)=w(\lambda)$, where $w(\lambda)$ denotes the image of the weight $\lambda$ under the Weyl group action, the character becomes
\begin{align}
\chi_{\lambda}&=\sum_{w \in W}w \bigg(\e^{\lambda} \prod_{\alpha \in \Delta^+}\frac{1}{1-\e^{-\alpha}}\bigg)\label{eq:pre-weyl-character}\ .
\end{align}
We can simplify further writing
\be 
\prod_{\alpha \in \Delta^+} (1-\e^{-\alpha})=\e^{-\rho} \prod_{\alpha \in \Delta^+} (\e^{\frac{\alpha}{2}}-\e^{-\frac{\alpha}{2}})\ , \label{eq:denominator-factoring}
\ee
with $\rho=\frac{1}{2} \sum_{\alpha \in \Delta^+} \alpha$ the Weyl vector. The Weyl group acts on this product by a sign:
\be 
w\bigg(\prod_{\alpha \in \Delta^+} (\e^{\frac{\alpha}{2}}-\e^{-\frac{\alpha}{2}})\bigg)=\epsilon(w)\prod_{\alpha \in \Delta^+} (\e^{\frac{\alpha}{2}}-\e^{-\frac{\alpha}{2}})\ , \label{eq:sign-action-denominator}
\ee
where $\epsilon:W \to \{\pm 1\}$ is the sign homomorphism. $W$ is generated by the simple reflections. We can thus express each $w \in W$ as a product $w=s_1 \cdots s_k$ of such simple reflections. Then $\epsilon(w)=(-1)^k$ is independent of the way we decompose $w$ as a product of simple reflections. In the example of $G=\SU(N)$, $W\cong S_N$ and $\epsilon$ coincides with the usual sign homomorphism on permutations.

Since $\epsilon$ is multiplicative, it suffices to verify \eqref{eq:sign-action-denominator} for a single simple reflection $s$, whose axis is along the simple root $\alpha_i$. The key observation is that $s$ sends $\alpha_i\mapsto -\alpha_i$ while permuting the remaining positive roots $\Delta^+\setminus\{\alpha_i\}$ among themselves. Therefore, exactly the factor $(\e^{\frac{\alpha_i}{2}}-\e^{-\frac{\alpha_i}{2}})$ changes sign, while the product over the remaining positive roots gets rearranged. Since $\epsilon(s)=-1$, this verifies \eqref{eq:sign-action-denominator}. Substituting \eqref{eq:denominator-factoring} into \eqref{eq:pre-weyl-character} and using \eqref{eq:sign-action-denominator}, we obtain the Weyl character formula \cite{weyl1925theorie1, weyl1926theorie2, weyl1926theorie3},
\be\label{eq:weyl-character}
\chi_{\lambda}=\frac{\sum_{w \in W} \epsilon(w)\,\e^{w(\lambda+\rho)}}{\prod_{\alpha \in \Delta^+}(\e^{\frac{\alpha}{2}}-\e^{-\frac{\alpha}{2}})}\ .
\ee
\paragraph{Other orbits.} We derived the character formula \eqref{eq:weyl-character} for generic orbits of the form $G/T$. When $\lambda$ lies on a wall of the Weyl chamber, i.e.\ $\langle\lambda,\alpha^\vee\rangle=0$ for some root $\alpha$, the stabilizer enhances and the orbit is of the form $G/H$ with $H=\exp \mathfrak{h}$ where $\mathfrak{h}$ is generated by $\mathfrak{t}$ as well as $E_\alpha$ and $F_\alpha$ for all roots with $\langle \lambda,\alpha^\vee \rangle=0$. This can also be diagnosed by noting that the KKS symplectic form on $G/T$ becomes degenerate in that case. The calculation \eqref{eq:kks-reduction} shows that it is trivial on the sphere $\mathrm{S}^2_\alpha$. In such a case, we have two options: (i) quantize the actual coadjoint orbit $\mathcal{O}_\lambda \cong G/H$ with its (non-degenerate) KKS form, or (ii) continue working with $G/T$ as in the derivation above, equipped with the degenerate pullback of $\omega_\lambda$. 
In this case, both give the same result. Indeed, we have the fibration $H/T \to G/T \to G/H$ with the symplectic form being trivial on $H/T$. Thus when quantizing $G/T$, wavefunctions are constant along $H/T$ and the resulting representation spaces are isomorphic. For this argument, it is important that $H/T$ is a compact space, so that the trivial line bundle on it only supports the constant section.

This means in particular that the Weyl character formula \eqref{eq:weyl-character} holds for all weights $\lambda$ including those with an extended stabilizer. For $\lambda=0$ the orbit is a point and $\chi_0=1$; the Weyl character formula then reduces to the Weyl denominator formula
\be
\sum_{w\in W}\epsilon(w)\,\e^{w(\rho)}=\prod_{\alpha\in\Delta^+}\bigl(\e^{\frac{\alpha}{2}}-\e^{-\frac{\alpha}{2}}\bigr)\ .
\ee

\section{Application to conformal field theory} \label{sec:application-to-cft}
\subsection{Virasoro algebra} \label{subsec:virasoro-algebra}
In the present Section, we will review the application of the coadjoint orbit procedure to the Virasoro group described in \cite{Witten:1987ty, Alekseev_Shatashvilli_PI}. This will lead to the Virasoro character formula for unitary representations. The vacuum representation was already partially treated in this way in \cite{Eberhardt:2022wlc}.

\paragraph{Definition.} The Virasoro group is an infinite-dimensional Lie group which can be defined as the central extension of the orientation preserving diffeomorphism group of the circle, $\mathrm{Vir}=\widehat{\DiffS}$. The Lie algebra $\mathfrak{vir}$ consists of a central extension of infinitesimal diffeomorphisms, i.e.\ vector fields. The standard Virasoro generators correspond to the basis 
\be 
L_n=i\,\e^{i n \varphi} \partial_\varphi+\frac{c}{24} \delta_{n,0}\ , \label{eq:virasoro-generators}
\ee
which, after central extension, satisfy the Virasoro algebra $[L_m,L_n]=(m-n) L_{m+n}+\frac{c}{12}m(m^2-1) \delta_{m+n,0}$ and the reality condition $L_n^\dag=L_{-n}$.
The dual of the Virasoro algebra can be realized as quadratic differentials, coupled to a central element $c$,
\be
(b(\varphi)\, \d\varphi^2 , c) \in \mathfrak{vir}^*\ ,
\ee
with $b(\varphi)$ a smooth periodic function of the angle $\varphi$. Quadratic differentials $b$ are naturally dual to vector fields $\alpha=\alpha(\varphi) \partial_\varphi$ via the pairing $\langle b,\alpha \rangle=\int_{\mathrm{S}^1} b \,\alpha$. The central element $c$ by definition does not transform under the coadjoint action and we may restrict the coadjoint orbit to the space of quadratic differentials.
Explicitly, the coadjoint action of a circle diffeomorphism $f$ is given by
\be
f \cdot (b(\varphi)\, \d\varphi^2)=\Big(b(f^{-1}(\varphi))(f^{-1})'(\varphi)^2-\frac{c}{24\pi}\{f^{-1},\varphi\}\Big)\d\varphi^2\ , \label{eq:virasoro-coadjoint-action}
\ee
where $\{f^{-1},\varphi\}$ denotes the Schwarzian derivative. We should mention that there are equivalent ways to write \eqref{eq:virasoro-coadjoint-action}, related by adding coboundaries to the Schwarzian derivative. Another choice we could make is to replace $\{f^{-1},\varphi\}$ by $\{f^{-1},\varphi\}+\frac{1}{2}((f^{-1})'(\varphi))^2-\frac{1}{2}$ which can be absorbed by shifting $b(\varphi) \to b(\varphi)+\frac{c}{48\pi}$ on both sides. With the inclusion of this shift, the central term vanishes for the $\PSL(2,\RR)$ subgroup of $\DiffS$. Indeed, that subgroup is realized by maps
\be 
\varphi \mapsto 2 \arctan\Big(\frac{\alpha \tan \varphi/2+\beta}{\gamma \tan \varphi/2+\delta}\Big)\ ,
\ee
on which the central term can be checked to vanish.
This shift corresponds to the familiar shift of the $L_0$ mode of the Virasoro algebra when mapping from the punctured plane to the cylinder. The convention \eqref{eq:virasoro-coadjoint-action} automatically incorporates the correct ground state energy into the characters and is thus preferable to work with.

\paragraph{Coadjoint orbits.} Among the coadjoint orbits of the Virasoro group, we will focus on those that contain a constant representative $b_0 \d \varphi^2 \in \mathfrak{vir}^*$. 
We will begin by analyzing the structure of such orbits by showing that the value of $b_0$ is unique. As a byproduct, this will also give a description of the stabilizer.

Suppose that there is a diffeomorphism $f \in \DiffS$ such that $b_0f'(\varphi)^2-\frac{c}{24\pi} \{f,\varphi\}$ is also constant, say $B_0$. We replace $f^{-1}$ with $f$ compared to \eqref{eq:virasoro-coadjoint-action} for convenience. Since $f'(\varphi)>0$, we can define $\psi(\varphi)=f'(\varphi)^{-1}$. Then $\{f,\varphi\}=-\frac{\psi''(\varphi)}{\psi(\varphi)}+\frac{\psi'(\varphi)^2}{2\psi(\varphi)^2}$, so the constraint after multiplication by $\psi'(\varphi) $ becomes
\be 
B_0\psi'(\varphi)=b_0 \psi(\varphi)^{-2}\psi'(\varphi)-\frac{c \psi'(\varphi)}{24\pi}\Big(-\frac{\psi''(\varphi)}{\psi(\varphi)}+\frac{\psi'(\varphi)^2}{2\psi(\varphi)^2}\Big)\ . \label{eq:differential-equation-1}
\ee
We can integrate both sides of the equation and get
\be 
B_0 \psi(\varphi)+A=-\frac{b_0}{\psi(\varphi)}+\frac{c\psi'(\varphi)^2}{48\pi \psi(\varphi)}\ , \label{eq:differential-equation-2}
\ee
where $A$ is the integration constant. Multiply by $\psi(\varphi)$ and take a derivative to obtain
\be
\psi'(\varphi)=0\quad \text{or}\quad -\frac{c}{24\pi} \psi''(\varphi)+2B_0 \psi(\varphi)+A =0\ . \label{eq:differential-equation-3}
\ee
The former case together with the periodicity condition $f(\varphi+2\pi)=f(\varphi)+2\pi$ leads to translations $f(\varphi)=\varphi+\text{const}$. The latter case is an inhomogeneous linear differential equation. Assuming $B_0 \ne 0$, the solution takes the form
\be 
\psi(\varphi)=-\frac{A}{2B_0}+C \e^{i \omega \varphi}+C^* \e^{-i \omega \varphi}\ , \quad \omega=\sqrt{-\frac{48\pi B_0}{c}}\ , \label{eq:psi-solution}
\ee
with $C$ the complex-valued integration constant. For $B_0=0$, the solution is incompatible with the periodicity properties of $f(\varphi)$. From \eqref{eq:differential-equation-2}, one can verify the relation 
\be 
A^2-4b_0B_0-16B_0^2 |C|^2=0\ . \label{eq:integration-constant-relation}
\ee
We can assume that $C \ne 0$, since otherwise the solution reduces to constant $\psi(\varphi)$. This solution has to respect the periodicity conditions of $f$, which implies that $\omega=k$ for $k \in \NN$, i.e.\ $B_0=-\frac{k^2 c}{48\pi}$. Lastly, we need to verify compatibility with $f(\varphi+2\pi)=f(\varphi)+2\pi$, which is the condition
\be 
2\pi=\int_0^{2\pi} \frac{\d \varphi}{\psi(\varphi)}=\oint \frac{\d z}{i z} \Big(-\frac{A}{2B_0}+C z+C^* z^{-1}\Big)^{-1}=2\pi \sqrt{\frac{B_0}{b_0}}\ .
\ee
We used the change of variables $z=\e^{i \omega \varphi}$ and the relation \eqref{eq:integration-constant-relation}. We thus learn that in all cases $b_0=B_0$. 

Thus, for generic values of $b_0$, the stabilizer group consists only of translations $\varphi \mapsto \varphi+\text{const}$, which gives rise to coadjoint orbits of the form $\DiffS /\mathrm{S}^1$. Since the Virasoro group contains the subgroup $\PSL(2,\RR)$ of M\"obius transformations, the coadjoint orbits of the Virasoro group will contain, as a submanifold, a coadjoint orbit of $\PSL(2,\RR)$ which are known to be hyperboloids \cite{Witten:1987ty}. Embedded in $\sl(2,\RR)^* \cong \RR^3$, with coordinates $(x,y,z)$, these submanifolds are given by $z^2-x^2-y^2=h^2$. Then, one can label generic Virasoro orbits by the radius $h$ of the hyperboloid they contain. 
These orbits do not quantize the value of $h$ or $c$. Indeed, since $\DiffS$ retracts onto $\mathrm{S}^1$, the quotient $\DiffS/\mathrm{S}^1$ is contractible and the quantization condition vacuous. 
Notice that for the line bundle to be positive, we have to assume that $h$ is sufficiently large, which is the geometric incarnation of the unitarity condition $h \ge 0$ visible on the level of the Virasoro algebra.

For the particular values $b_0=-\frac{k^2 c}{48 \pi}$ with $k \in \NN$, there are three Virasoro generators $L_{-k}$, $L_0$ and $L_k$ defined by \eqref{eq:virasoro-generators} that leave $b_0$ invariant. These are the infinitesimal actions corresponding to the solution \eqref{eq:psi-solution} above. Therefore, the stabilizer is the group generated by those Lie algebra elements, $H_{-\frac{k^2c}{48\pi}}=\PSL^{(k)}(2,\RR)$. We then conclude that for such values of $b_0$, there exist coadjoint orbits with a more exotic geometry,
\be\label{eq:virasoro-exotic-orbits}
    \mathcal{O}_{-\frac{k^2c}{48\pi}}\cong \DiffS/\PSL^{(k)}(2,\RR)\ .
\ee
The criterion \eqref{eq:witten-criterion} implies that only the exotic orbit with $k=1$ can be endowed with an appropriate K\"ahler structure, see \cite{Witten:1987ty}. For the exotic orbits with $k \ge 2$, it does not work to define $\mathcal{K}^+=\mathrm{span}\{ L_{n},\, n \in \NN \setminus \{k\}\}$. Indeed, for $0<n<k$, $[L_{-k},L_{n}]=-(n+k) L_{n-k}$ is not contained in $\mathfrak{h} \oplus \mathcal{K}^+$, therefore violating the condition \eqref{eq:hkp-criterion}.

The holomorphic tangent directions at the identity coset, spanned by the annihilation modes, are given by
\be
\mathcal K_{\gen}^+ =\mathrm{span}\{L_{n}, \, n\geq 1\}\ ,
\ee
for generic coadjoint orbits $\DiffS/\mathrm{S}^1$, and
\be
\mathcal K_{\vac}^+ =\mathrm{span}\{L_{n},\,  n\geq 2\}\ ,
\ee
for the $\DiffS/\PSL(2,\RR)$ vacuum orbit. 

\paragraph{Geometric realization of coadjoint orbits.} The vacuum orbit $\DiffS/\PSL(2,\RR)$ embeds in the universal Teichm\"uller space and can be thought of as describing complex structures on the hyperbolic disk. While a disk with finite boundaries has a unique complex structure thanks to the Riemann mapping theorem, a disk with an asymptotic boundary also contains the data of how the boundary of the disk is glued to infinity. The gluing data naturally defines an element of $\DiffS$. The disk has automorphism group $\PSL(2,\RR)$, which naturally leads to the coset $\DiffS/\PSL(2,\RR)$. Similarly, the generic orbit $\DiffS/\mathrm{S}^1$ can be realized as the space of complex structures on the punctured hyperbolic disk. The additional puncture breaks the automorphism group to the subgroup $\mathrm{S}^1 \subset \PSL(2,\RR)$. This geometric picture naturally connects to 3d quantum gravity with negative cosmological constant, see \cite{Brown:1986nw, Eberhardt:2022wlc}.

\paragraph{Formal localization principle.} At this point, one could na\"ively write down an integral formula analogous to \eqref{eq:character-equivariant-index-theorem}, for the Virasoro character. Since the coadjoint orbit is infinite-dimensional, the initial index theorem is not well-defined. As suggested in \cite{Alekseev:2020jja}, one should take the localized formula \eqref{eq:localized-character} as the \emph{definition} of the integral \eqref{eq:character-equivariant-index-theorem} in the infinite-dimensional setting. As we shall see, it remains well-defined and without subtleties. With this in mind, we can go ahead and evaluate the ingredients of the localization formula. We will localize with respect to rigid rotation generated by $L_0$.

\paragraph{Fixed points.} For a generic point of the coadjoint orbit, the coadjoint action of a rigid rotation reduces to
\be
t \cdot(b(\varphi) \d\varphi ^2)=b(\varphi+t) \d\varphi ^2\ ,
\ee
where $t$ shifts the angle. Clearly, the only fixed points of this action are the points with a constant quadratic part and as stated before, there is a unique point of this type in each coadjoint orbit. Therefore, there is a unique fixed point.

\paragraph{Evaluation of the index theorem.} The momentum map of the $\U(1)$ action was given in \cite{Alekseev:2020jja} by 
\be \label{eq:momentum-map-virasoro}
\mu_b(t)=t\int_{\mathrm{S}^1} b(\varphi)\, \d\varphi\ ,\quad t \in \u(1) \ .
\ee
Hence, for constant $b$,
\be 
\mu_{b_0}(t)=2\pi t b_0\ .
\ee
To evaluate the contribution of the Todd and Euler classes as in \eqref{eq:equivariant-todd-euler-ratio}, we are left to compute the spectrum of the infinitesimal action at the fixed point. This linear operator acts on the tangent space at this point, which can be identified with the space of infinitesimal quadratic differentials of the form
\be
f(\varphi)=\sum_{n \in \ZZ\setminus\{0\}} f_n\, \e^{in\varphi} \d \varphi^2
\ee
for generic orbits with $f_n=f_{-n}^*$. For the vacuum orbit, the stabilizer $\PSL(2,\RR)$ eliminates the modes $f_{\pm 1}$ as well, so the sum is further restricted to $|n|\geq 2$. The operator $t L_0=i t\partial_\varphi$ (cf. \eqref{eq:virasoro-generators}) with respect to which we are localizing acts on these tangent vectors. We are interested in the spectrum when restricted to the holomorphic tangent direction $\mathcal{K}^+$. Since $[L_0,L_n]=-n L_n$, this is $-t \NN$ for the generic orbit and $-t \NN_{\ge 2}$ for the vacuum orbit, respectively. 

Therefore, \eqref{eq:localized-character} becomes for a generic orbit
\be 
\chi_{h}(t)=\frac{\e^{2\pi t b_0}}{\prod_{m=1}^\infty (1-\e^{m t})}=\frac{q^{h-\frac{c}{24}}}{\prod_{m=1}^{\infty} (1- q^{m})}\ . \label{eq:generic-Virasoro-character}
\ee
We identified $q=\e^{t}$ and $2 \pi b_0=h-\frac{c}{24}$ to pass to the standard form of the character. This identification of $b_0$ with the conformal weight is consistent with $b_0=-\frac{c}{48\pi}$ describing the vacuum orbit. In that case, the character reads
\begin{align}
    \chi_{\vac}(t)=\frac{q^{-\frac{c}{24}}}{\prod_{m=2}^{\infty} (1- q^m)}\ . \label{eq:vacuum-Virasoro-character}
\end{align}
Throughout, the infinite products may be understood as formal expressions of $q$ or as analytic expressions if we consider $|q|<1$ in the complex disk.
\subsection{Affine Kac--Moody algebras} \label{subsec:affine-kac-moody}
We now turn our attention to the representation theory of affine Kac--Moody algebras that appear as the symmetry algebras of WZW models. We will focus in our discussion mostly on the case of the $\su(2)$ Kac--Moody algebra. These infinite-dimensional Lie algebras can be realized as the Lie algebras of centrally extended loop groups $\widehat{LG}$, \cite{loopgroup_Pressley_Segal}. The large-$k$ limit of the following discussion can also be found in \cite{Alekseev:2020jja}.

\paragraph{Definition.} An element of the loop group $LG$ is a map $g:\mathrm{S}^1 \to G$. Its Lie algebra is naturally realized by maps $X:\mathrm{S}^1 \to \mathfrak{g}$. Assuming that $\mathfrak{g}$ is simple, the dual of the Lie algebra can be identified as Lie algebra-valued 1-forms $A=A(\varphi) \,\d \varphi$ via the pairing 
\be 
\langle A,X\rangle=\int_{\mathrm{S}^1} \tr(AX)\ , \label{eq:loop-group-pairing}
\ee 
where the trace is taken in a faithful representation and proportional to the Killing form. We think of $A$ as gauge fields. The central extension appends a number $k \in \RR$ called the level. As in the Virasoro case, the central extension only affects the action of the loop group on the gauge field.

Adopting the notation of \cite{Alekseev:2020jja}, the coadjoint action of $g(\varphi)\in LG$ on a dual Lie algebra element $A \in \Omega^1(\mathrm{S}^1, \mathfrak{g})$ is given by
\be\label{eq:coadjoint-action-loop-group}
g(\varphi) \cdot (A,k)= \left(gA\, g^{-1} -i k(\partial_\varphi g\, g^{-1}) \d \varphi,\, k\right)\ .
\ee
We take $\varphi$ to be $2\pi$-periodic as above and we use the physics convention $g=e^{iX}$, where $X$ is a Lie algebra element. This is precisely the action of gauge transformations on the gauge field $\frac{1}{k} A$ on the circle. For compact and simple $G$, the only gauge-invariant operator one can build out of $A$ is the Wilson line, i.e.\ the holonomy of $k^{-1}A$ around the circle, which defines a conjugacy class $\mathcal{C}$ in $G$. Therefore, the coadjoint orbits $\mathcal O_{\mathcal C}$ are besides the value of $k$ naturally labelled by conjugacy classes $\mathcal C \subset G$ specifying the holonomy of the gauge field around the circle \cite{Alekseev:2020jja}, 
\be
\mathcal O_{\mathcal C}=\{A \in \Omega^1(\mathrm{S}^1,\mathfrak{g}) \,| \,\text{Hol}(\mathrm{S}^1, k^{-1}A)\in \mathcal{C}\}\ . \label{eq:affine-kac-moody-orbit} 
\ee
These coadjoint orbits arise naturally as the phase space of 3d Chern--Simons theory on the manifold $\DD^* \times \RR$, whose time slices are punctured disks. Then, the conjugacy class prescribes the holonomy of the flat connections around the puncture \cite{Elitzur:1989nr}. 

Every orbit contains a $\varphi$-independent representative $A$. One can furthermore conjugate this constant representative to lie in the Cartan subalgebra $\mathfrak{t}$. Its stabilizer consists of constant maps $g:\mathrm{S}^1 \to T$ and can therefore be identified with the Cartan torus $T \subset G$. Thus, generic orbits are of the type $LG/T$, where $T\subset G$ is the maximal torus. Similarly to the finite-dimensional case discussed in Section~\ref{subsec:weyl-character-formula}, the stabilizer is enlarged in some cases. For example, when the orbit passes through the zero connection, the stabilizer consists of all constant maps into $G$ and the coadjoint orbit is of the form $LG/G$, sometimes referred to as the vacuum orbit. There are in general also many intermediate cases. 

\paragraph{K\"ahler structure and quantization.} In order to simplify the discussion, we will consider the case $G=\SU(2)$, whose maximal torus is $\U(1)$. The general case is treated at the end of this subsection. In this setting, we can explicitly define an adequate complex structure for the orbits: For the generic ones, we can take the holomorphic tangent directions at the identity to be
\be \label{eq:complex-structure-kac-moody}
\mathcal{K}_{\gen}^+=\mathrm{span}\{J_0^+,\; J_{m}^-,\; J_{m}^+,\; J_{m}^3\,:\, m\geq 1\}\ ,
\ee
which satisfies the criterion \eqref{eq:witten-criterion}.
For the vacuum orbit, the stabilizer includes all three zero-modes $\{J_0^+, J_0^-, J_0^3\}$, so the complex structure is given by
\be\label{eq:complex-structure-kac-moody-vacuum}
\mathcal{K}_{\vac}^+=\mathrm{span}\{J_{m}^+,\; J_{m}^-,\; J_{m}^3\,:\, m\geq 1\}\ .
\ee
As explained below equation~\eqref{eq:affine-kac-moody-orbit}, each orbit possesses a constant representative of the form $\xi d\varphi$ with $\xi \in \mathfrak{t}\subset \su(2)$.  In particular, identifying $\su(2)\cong \su(2)^*$ by means of the trace,  the coadjoint orbit of $\SU(2)$ obtained by acting with constant maps $g \in \SU(2) \subset L\SU(2)$ is a submanifold of the affine coadjoint orbit.
Requiring the periods of the KKS form to be integer on this submanifold imposes that the $\SU(2)$ weight is integer. We will label the representation in terms of the $\SU(2)$ spin $j$ and therefore label coadjoint orbits by $\mathcal O_{j,k}$ with $j \in \frac{1}{2}\ZZ$ and $k$ the level.
Requiring the KKS to live in integer cohomology also quantizes $k$. A simple intuitive reason to expect this is the fact $\text{rank} \, H^2(L\SU(2)/\U(1),\ZZ)=2$ \cite{10.1007/BFb0084589} and thus there are two quantization conditions, one for $j$ and one for $k$.
We will see a more quantitative statement of this quantization below.

\paragraph{Momentum maps.} We will localize with respect to two $\U(1)$ actions. One rotates the loop and can be thought of as the Hamiltonian in CFT. The other is inherited from the action of the Cartan torus $T \subset \SU(2)$ as in the finite-dimensional case discussed in Section~\ref{subsec:weyl-character-formula}. They act as
\be\label{eq:loop-group-toric-action}
(g,t) \cdot A(\varphi)\, \d \varphi= g\, A(\varphi + t)\, g^{-1}\, \d \varphi\ ,
\ee
where $g \in \U(1)$ is an element of the maximal torus $T \subset \SU(2)$, while $t \in \RR$ time translates. The momentum map $\mu: \mathcal O_{\mathcal C} \to \u(1)^* \oplus \u(1)^*$ associated with this Hamiltonian action is given by \cite{Alekseev:2020jja, Harada2005ConnectivityPO} 
\be\label{eq:momentum-map-kac-moody}
\mu(A)(s,t)= \frac{1}{4\pi}\int_{\mathrm{S}^1} \tr\!\big(A(\varphi)\diag(s,-s)\big)\, \d \varphi+\frac{t}{4\pi k}\int_{\mathrm{S}^1} \tr\big(A(\varphi)^2\big)\, \d\varphi\ .
\ee
Here $t \in \RR$, and the hermitian representative of the Cartan element is $\diag(s,-s) \in \u(1) \subset \su(2) \subset \widehat{\su(2)}$. Notice in particular that the first term is simply the pairing of $A$ with $\diag(s,-s)$ via \eqref{eq:loop-group-pairing}. Via the bilinear form, this is just the inclusion map of the coadjoint orbit into its dual Lie algebra as in the general case, see \eqref{eq:momentum-map}. The second term in \eqref{eq:momentum-map-kac-moody} is the momentum map for the translations along the circle. It has the standard bilinear structure that one also encounters in the Sugawara construction of the stress tensor.

\paragraph{Fixed-point set.} To study the fixed-point set of the toric action, let us pick the coadjoint orbit of $\widehat{L\SU(2)}$ whose spherical submanifold has radius $j\in \frac{1}{2} \NN_0$, i.e.\ 
\be 
A=\diag(j,-j)\,\d \varphi\ , \qquad 
\mathcal{C}_j=
\diag(\e^{2\pi i\phi}, \e^{-2\pi i\phi})\ ,\quad \text{where }\,\phi=\frac{j}{k}\ , \label{eq:holonomy-j-conjugacy-class}
\ee
with $\mathcal{C}_j=\text{Hol}(\mathrm{S}^1,k^{-1}A)$ the associated holonomy. Only the choices $j \in \{0,\frac{1}{2},\dots,\frac{k}{2}\}$ lead to different conjugacy classes and thus there are only finitely many representations.

The fixed points of the action \eqref{eq:loop-group-toric-action} on the orbit $\mathcal O_{j,k}$ are given by
\be \label{eq:fixed-points-kac-moody}
A=\begin{pmatrix}
\pm j +n k & 0 \\ 0 & \mp  j-nk
\end{pmatrix} \d \varphi\ , \quad n \in \ZZ\ .
\ee
The two signs generalize the north and south pole of the finite-dimensional case. The associated holonomy is conjugate to \eqref{eq:holonomy-j-conjugacy-class} for any $n \in \ZZ$ and the integer shifts can be obtained from the action of $g(\varphi)=\diag(\e^{i n \varphi},\e^{-i n \varphi})$ on the basic representative \eqref{eq:holonomy-j-conjugacy-class}. Together with the finite-dimensional Weyl group $W \cong \ZZ_2$ that exchanges north and south pole, this defines the affine Weyl group $\widehat{W}=W \ltimes \ZZ$.

Finally, we must compute the spectrum of the infinitesimal action at each fixed point and extract the charges of both $\u(1)$ actions. We can parametrize the perturbation at $A=\diag(j,-j) \d \varphi$ as
\be 
A(\varphi)=\sum_{n \in \ZZ } \e^{i n \varphi}((j \delta_{n,0}+A_{n,0}) t^3+A_{n,+} t^++A_{n,-}t^-)\ ,
\ee
with $t^3$, $t^+$ and $t^-$ the standard $\su(2)$ generators and $A_{0,0}=0$ since it corresponds to the stabilizer of the coadjoint orbit. The charge under the first momentum map in \eqref{eq:momentum-map-kac-moody} for the direction given by $A_{n,a}$ is $a \in \{-1,0,1\}$, while the charge under the second momentum map is $n \in \ZZ$. Thus the joint spectrum under the full momentum map is $\lambda \in \{(a,n)\}_{a \in \{-1,0,1\},n \in \ZZ} \setminus \{(0,0)\}$.
\paragraph{The complex structure and the $\widehat{A}$-genus.} It remains to pick a complex structure, i.e.\ a subset of the spectrum consistent with the conditions \eqref{eq:witten-criterion}. Contrary to the previous cases, there are many possible choices. In particular, we can pick
\be 
\{(1,n)\}_{n \le -w-1}\cup \{(0,n)\}_{n \le -1} \cup \{(-1,n)\}_{n \le w}
\ee
for any choice of $w \in \ZZ$. We only have the freedom to make this choice at one fixed point, which will dictate the corresponding choice at all other fixed points. A convenient way to package this is as follows. Ultimately, we will need to evaluate the equivariant Todd class as in \eqref{eq:equivariant-todd-euler-ratio}. As already in the finite-dimensional case \eqref{eq:sign-action-denominator}, a simpler quantity is the following symmetrized product,
\be 
\frac{\td_p(T \mathcal{O}_\lambda)}{e_p(T \mathcal{O}_\lambda)}=\prod_{i \ge 1} \frac{1}{1-\e^{-x_i}}=\e^{\frac{1}{2}\sum_{i\geq1} x_i}\prod_{i\geq1} \frac{1}{\e^{\frac{1}{2} x_i}-\e^{-\frac{1}{2} x_i}}\ . \label{eq:passage-to-a-roof}
\ee
This corresponds to the rewriting of the integrand of the index theorem \eqref{eq:character-equivariant-index-theorem}
\be 
\int_{\mathcal{O}_\lambda} \big(\widehat{A}^G(T \mathcal{O}_\lambda) \, \e^{\omega^G-\frac{1}{2}c_1(\mathcal{K}_{\mathcal{O}})} \big)(X)\ .
\ee
Here, $\mathcal{K}_{\mathcal{O}}=\bigwedge^\text{top} T^* \mathcal{O}_\lambda$ is the canonical line bundle on the coadjoint orbit and $\widehat{A}$ the A-roof genus. The main reason for this rewriting is that $\widehat{A}$ is the same for every fixed point, up to signs, since terms in the product \eqref{eq:passage-to-a-roof} will just get permuted. The subtraction of $\frac{1}{2}c_1(\mathcal{K}_{\mathcal{O}})$ will then effectively modify the line bundle under consideration by some fixed amount. Thus, the new line bundle for the orbit $\mathcal{O}_{j,k}$ is $\mathcal{L}_{j,k} \otimes \mathcal{K}_{\mathcal{O}}^{-\frac{1}{2}}$. Every line bundle on $\mathcal{O}_{j,k}$ is of the form $\mathcal{L}_{j',k'}$ for some value of $j'$ and $k'$. We show in Appendix~\ref{subapp:affine-ghosts} that for the canonical line bundle $\mathcal{K}_{\mathcal{O}} \cong \mathcal{L}_{j=-1,k=-4}$. The computation proceeds by noting that one can construct a section of this line bundle from a $\mathfrak{b}\mathfrak{c}$ ghost system (that appears when one tries to gauge the affine currents). The spin and level can be read off from the free-field representation of the current algebra.

This simply induces a shift in both $j$ and $k$ yielding the following expression for the character
\be
\chi_{j,k}=\int_{\mathcal{O}_{j,k}} \widehat{A}(T\mathcal O_{j,k})\, \e^{c_1(\mathcal L_{j+1/2,k+2})}\ . \label{eq:line-bundle-shift}
\ee
\paragraph{Evaluation.} We can finally evaluate the remaining expression. The denominator at each fixed point can be evaluated with the aid of $\zeta$-function regularization,
\begin{align}
        \prod_\lambda (\e^{\frac{1}{2}\lambda}-\e^{-\frac{1}{2}\lambda})&= \e^{-\frac{3}{2}\zeta(-1) t} \big(\e^{\frac{1}{2}s}-\e^{-\frac{1}{2}s}\big) \prod_{m=1}^\infty \big(1-\e^{mt}\big)\big(1-\e^{mt+s}\big)\big(1-\e^{mt-s}\big)\\
        &=q^{\frac{1}{8}} (y^{\frac{1}{2}}-y^{-\frac{1}{2}}) \prod_{m=1}^\infty (1-q^m)(1-y\, q^m)(1-y^{-1}\, q^m)\ , \label{eq:kac-denominator}
\end{align}
where $q=\e^{t}$ and $y=\e^{s}$. Together with the sign $\varepsilon$ of $\widehat{A}$ at the fixed points that is the same as for the finite-dimensional case, we obtain the numerator 
\be 
\sum_{A \text{ fixed}} \varepsilon(A) \e^{\mu(A)(s,t)} = \sum_{\sigma=\pm 1} \sum_{n \in \ZZ} \sigma y^{ J}q^{\frac{1}{k+2} J^2}\big|_{J=\sigma(j+\frac{1}{2})+(k+2)n}\ . \label{eq:su2-sum-over-fixed-points}
\ee
The sum over $\sigma$ and $n$ parametrizes the fixed-point set \eqref{eq:fixed-points-kac-moody} with $\sigma$ representing the sign. $J$ stands for the eigenvalues of the matrix, while keeping the shift in \eqref{eq:line-bundle-shift} in mind. The term $y^J$ is the result of evaluating the first term in \eqref{eq:momentum-map-kac-moody} while $q^{\frac{1}{k} J^2}$ originates from the second term in \eqref{eq:momentum-map-kac-moody}, where the shift $k\to k+2$ from \eqref{eq:line-bundle-shift} has to be included.

When fully assembled, we recover the Weyl--Kac character formula 
\cite{kac1974infinite, Murthy:2025ioh}
\be
\chi_{j,k}(s,t)=\frac{q^{\frac{j(j+1)}{k+2}-\frac{k}{8(k+2)}}\sum_{n \in \ZZ}q^{(k+2)n^2+n(2j+1)}\big( y^{j+\frac{1}{2}+(k+2)n}-y^{-(j+\frac{1}{2}+(k+2)n)}\big)}{(y^{\frac{1}{2}}-y^{-\frac{1}{2}}) \prod_{m=1}^\infty (1-q^m)(1-y \,q^m)(1-y^{-1}\,q^m)}\ .
\ee
The prefactor $q^{\frac{j(j+1)}{k+2}}$ reflects the conformal weight of the primary state of spin $j$, while the shift $-\frac{k}{8(k+2)}$ in the exponent corresponds to the usual $-\frac{c}{24}$ shift in CFT, where $c=\frac{3k}{k+2}$ is the central charge of the $\widehat{\su(2)}$ current algebra at level $k$.

\paragraph{General $\widehat{\mathfrak{g}}_k$.} The above computation generalizes straightforwardly to an arbitrary affine Kac--Moody algebra $\widehat{\mathfrak{g}}_k$ at level $k$, where $\mathfrak{g}$ is the Lie algebra of a simple, simply-connected compact group $G$ of rank $r$. The generic coadjoint orbits of the centrally extended loop group $\widehat{LG}$ are of the form $LG/T$, where $T$ is now the rank-$r$ maximal torus of $G$, and the toric action is by $T \times \U(1)$, with rotation fugacity $q=\e^{t}$; we keep the Cartan dependence formal, writing $\e^{\alpha}$ for the character associated with a root $\alpha \in \mathfrak{t}^*$. The fixed points of this action are labeled by pairs $(w,\gamma) \in \widehat{W}=W \ltimes Q^\vee$, where $W$ is the Weyl group of $\mathfrak{g}$ and $Q^\vee$ its coroot lattice. For $\su(2)$, this reduces to the $\ZZ_2$ sign and the integer $n$ in \eqref{eq:fixed-points-kac-moody}. The spectrum of the infinitesimal toric action at each fixed point now consists of the eigenvalues $\{(m,\alpha)\}_{m \in \ZZ, \, \alpha \in \Delta}$ from the root directions, together with $\{(m,0)\}_{m \in \ZZ\setminus\{0\}}$ with multiplicity $r$ from the Cartan directions, where $\Delta$ denotes the set of roots of $\mathfrak{g}$. The canonical line bundle is $\mathcal{K}_{\mathcal{O}} \cong \mathcal{L}_{-2\rho, -2h^\vee}$ as shown in Appendix~\ref{subapp:affine-ghosts}, inducing the shifts $\lambda \to \lambda + \rho$ and $k \to k + h^\vee$, where $\rho$ is the Weyl vector. Assembling all ingredients and $\zeta$-regularizing the infinite products, the localized equivariant index theorem yields the general Weyl--Kac character formula
\be \label{eq:kac-weyl-general}
\chi_{\lambda,k}(t)=\frac{\sum_{w \in W}\epsilon(w)\,\Theta_{w(\lambda+\rho),k+h^\vee}(t)}{q^{\frac{1}{24}\dim \mathfrak{g}}\prod_{\alpha \in \Delta^+}\left(\e^{\alpha/2}-\e^{-\alpha/2}\right)\prod_{m=1}^\infty (1-q^m)^r\prod_{\alpha \in \Delta}(1-\e^\alpha q^m)}\ ,
\ee
where $\epsilon(w)=\det(w)$ is the sign of the Weyl group element, $\Delta^+$ denotes the set of positive roots, $\lambda$ is the highest weight of the representation, and we have defined the theta functions of the affine root lattice,
\be
\Theta_{\mu,K}(t)=\sum_{\gamma \in Q^\vee} \e^{\mu+K\gamma}\, q^{\frac{1}{2K}|\mu+K\gamma|^2} \ .
\ee
For $\mathfrak{g}=\su(2)$, one has $r=1$, $\Delta^+=\{\alpha\}$ with $\e^{\alpha}(s)=y$, and $Q^\vee=\ZZ$, and one readily recovers the character formula derived above.
As for the finite-dimensional case, validity of the character formulas extends to the case with extended stabilizers. 

\subsection{Super Virasoro algebra} \label{subsec:super-virasoro-algebra}
In the following sections, we will repeat the study of coadjoint orbits and the corresponding localized characters for several supersymmetric extensions of the Virasoro algebra. At the Lie group level, this amounts to examining conformal superdiffeomorphisms of supercircles $\mathrm{S}^{1|\mathcal{N}}$. We will discuss the cases where $\mathcal{N}=1$, $2$ and $4$.
\subsubsection{\texorpdfstring{$\mathcal{N}=1$}{N=1} Super Virasoro algebra}
\paragraph{Definition.} The first supersymmetric extension of the Virasoro algebra, $\mathfrak{vir}_{\mathcal{N}=1}$, is generated by the following modes:
\begin{itemize}
    \item The usual Virasoro generators $L_n$, $n \in \ZZ$,
    \item Fermionic supercurrents $G_r$, where $r \in \ZZ + \delta$. Here $\delta=1/2$ in the Neveu--Schwarz (NS) sector, and $\delta=0$ in the Ramond (R) sector.
\end{itemize}
Geometrically, $\mathfrak{vir}_{\mathcal{N}=1}$ is a central extension of the Lie algebra of superconformal diffeomorphisms, which we will define shortly. General superdiffeomorphisms, $\mathrm{Diff}\,\mathrm{S}^{1|1}$, feature, besides the usual bosonic angular coordinate $\varphi$, an additional Grassmann-odd coordinate $\theta$, representing a fermionic direction. These superdiffeomorphisms can be expanded as a Taylor series in the Grassmann variable as follows,
\be
    \Phi(\varphi, \theta)=\Phi_1(\varphi)+\theta \Phi_2(\varphi)\ ,\quad \Theta(\varphi,\theta)=\Theta_1(\varphi)+\theta\Theta_2(\varphi)
\ee
since $\theta^2=0$. Here, $(\Phi,\Theta)$ describes the image of the coordinates $(\varphi,\theta)$. Consequently $\Phi_1$ and $\Theta_2$ are Grassmann even, while $\Phi_2$ and $\Theta_1$ are Grassmann odd.
\paragraph{$\mathcal{N}=1$ Superconformal diffeomorphisms.}
In order to obtain the fully-fledged $\mathcal{N}=1$ super Virasoro group, we need to restrict the group of superdiffeomorphisms to its superconformal subgroup, $\widehat{\Diff_{\mathrm{sc}}\,\mathrm{S}^{1|1}}$. Superconformal diffeomorphisms are those that preserve the $(0|1)$ distribution generated by the super derivative $D=\partial_\theta + \theta \partial_{\varphi}$ \cite{Voronov:1987xf, Witten:2012ga}. This constraint reduces to the condition \cite{Fu:2016vas, Cotler:2018zff}
\be\label{eq:n1-superconformal-condition}
D \Phi= \Theta D \Theta\ .
\ee
Equivalently, superconformal diffeomorphisms are contactomorphisms with respect to the contact form
\be
\alpha=\d\varphi -\theta \,\d\theta\ ,
\ee
whose kernel is precisely spanned by $D$ \cite{Witten:2012ga, Michel2007OnTP}. 

This allows us to determine $\Phi_2$ and $\Theta_2$ with respect to $\Phi_1$ and $\Theta_1$, up to a choice of sign, fixed by the spin structure. Explicitly, one can write 
\be
\Phi_2=\pm\Theta_1 \sqrt{\partial_\varphi \Phi_1}\, , \quad \Theta_2=\pm\sqrt{\partial_\varphi\Phi_1+\Theta_1 \partial_\varphi \Theta_1}\ .
\ee
Another way to repackage the information is to consider the superconformal vector field $\mathcal{X}$ generating $(\Phi,\Theta)$. The constraint \eqref{eq:n1-superconformal-condition} can be solved by writing
\be 
\mathcal{X}_F=\Big(F-\frac{1}{2}\theta DF\Big)\partial_\varphi+\frac{
1}{2} D F \partial_\theta\ , \label{eq:n1-superconformal-vector-field}
\ee
where $F$ is an arbitrary superfunction. Thus, it is useful to parametrize superconformal diffeomorphisms by $F$.

The $\mathcal{N}=1$ super Virasoro algebra can then be taken to be a central extension of the commutator $[\mathcal{X}_F,\mathcal{X}_{\tilde{F}}]$. This leads to \cite{Delius:1990pt,Cotler:2018zff}
\be
[F,\tilde{F}] = F \partial \tilde F - \tilde F \partial F + \frac{1}{2} D F D \tilde F 
- \frac{c}{48\pi} \int \d\varphi \, \d\theta \, \big( F \partial^2 D \tilde F - \tilde F \partial^2 D F \big)\ . \label{eq:n1-supercommutator}
\ee
The last term is the $\mathcal{N}=1$ supersymmetric analogue of the Gelfand--Fuchs cocycle \cite{Delius:1990pt}. When expanding
\be 
F=i \sum_n L_n \e^{i n \varphi}+2 \e^{\frac{\pi i}{4}}\,\theta\sum_r G_r \e^{i r \varphi} \label{eq:f-fourier-series}
\ee 
as a Fourier series, the bracket \eqref{eq:n1-supercommutator} reproduces the $\mathcal{N}=1$ super Virasoro algebra relations
\begin{subequations}
    \begin{align}
        [L_m,L_n]&=(m-n)L_{m+n}+\frac{c}{12} m^3 \delta_{m+n,0}\ , \\
        [L_m,G_r]&= \left(\frac{m}{2}-r\right) G_{m+r}\ , \\
        \{G_r,G_s\}&=2 L_{r+s}+\frac{c}{3} r^2\, \delta_{r+s,0}\ . 
    \end{align} \label{eq:n1-virasoro-algebra}%
\end{subequations}
One can restore the more standard form of the central extension by shifting $L_0 \mapsto L_0+\frac{c}{24}$.

\paragraph{Coadjoint orbits.}
To describe the geometry of the coadjoint orbits (see e.g.\ \cite{Cotler:2018zff}), it is enough to find the stabilizer group $H_\lambda$ of the coadjoint element defining the orbit. The central $\U(1)$ acts trivially, so one obtains orbits of the form
\be
\mathcal{O}_\lambda \cong \mathrm{Diff}_{\mathrm{sc}}\,\mathrm{S}^{1|1}/H_\lambda\ .
\ee
The infinitesimal coadjoint action of a superconformal vector field $\mathcal 
{X}_F\in \mathfrak{vir}_{\mathcal{N}=1}$ on a dual element $B \in \mathfrak{vir}_{\mathcal{N}=1}^*$ coupled to the central element $c$ is given by \cite{Cotler:2018zff},
\be
    \delta_F B=F\partial B+\frac{3}{2}B \partial F + \frac{1}{2}DF\, DB-\frac{c}{24 \pi}\partial^2 DF \ . \label{eq:n1-coadjoint-orbit-action}
\ee
We can obtain the generators of $H$ by finding all superconformal vector fields $\mathcal X_F$ such that $\delta_F B=0$. As in the bosonic case, we will be probing coadjoint orbits that pass through a constant dual element $B=\theta b_0$. Expanding $F$ in a Fourier series as in \eqref{eq:f-fourier-series}, we get the following equations,
\begin{subequations}
\begin{align}
        &\sum_{r \in \ZZ+\delta}G_r\, \e^{ir\varphi} \left( \frac{1}{2}b_0+\frac{c}{24 \pi}r^2\right)=0\ , \\
        &\sum_{n \in \ZZ \setminus \{0\}}L_n\,\e^{in\varphi}\left( 2b_0 n +\frac{c}{24 \pi}n^3\right)=0\ .
\end{align}
\end{subequations}
For generic values of $b_0$, the conditions above are satisfied only if all Fourier coefficients vanish, i.e.\ $L_n=0$ and $G_r=0$ for all $n\ne 0$ and $r \in \ZZ+\delta$. In this case, the geometry of the coadjoint orbit is $\mathrm{Diff}_{\mathrm{sc}}\,\mathrm{S}^{1|1}/\mathrm{S}^1$, with the stabilizer $\mathrm{S}^1$ generated by $L_0$.
For the special values
\begin{align}
    b_0=-\frac{c}{12 \pi}r^2  \label{eq:n1-exceptional-b0-1}
\end{align}
with $r \in \NN_0+\delta$,
we find that the corresponding stabilizer includes the generators $G_{r}$ and $G_{-r}$. Similarly, for
\be
    b_0=-\frac{c}{48\pi}n^2 \label{eq:n1-exceptional-b0-2}
\ee
with $n \in \NN$, the stabilizer includes the generators $L_n$ and $L_{-n}$.

\paragraph{K\"ahler structures.}
Recall that, denoting by $\mathcal{K}$ the complement of the stabilizer's Lie algebra $\mathfrak{h}=\mathrm{Lie}(H)$, a K\"ahler structure splits $\mathcal{K}$ into holomorphic and anti-holomorphic parts,
\be
\mathcal{K}=\mathcal{K}^+\oplus\mathcal{K}^-\ .
\ee
In the case of generic coadjoint orbits, $\mathfrak{h}=\mathrm{span}\{L_0\}$. For this reason, in the Ramond-sector we need to allocate $G_0$ to either $\mathcal{K}^+$ or $\mathcal{K}^-$. Without loss of generality, choosing $G_0 \in \mathcal{K}^+$ implies $G_0^\dagger \in \mathcal{K}^-$; but $G_0$ is Hermitian, so $G_0=G_0^\dagger \in \mathcal{K}^+\cap\mathcal{K}^-=\{0\}$, a contradiction. Therefore, the coadjoint orbit cannot be endowed with an appropriate K\"ahler structure.\par
We do not encounter such an issue in the NS-sector, since the fermionic generators are indexed by half-integers. In this case, we can simply define
\be
\mathcal{K}_{\gen}^+=\mathrm{span}\left\{L_{n}, G_{r}\,:\,n\geq 1,\; r \geq \frac{1}{2}\right\}\ , \label{eq:n1-virasoro-generic-kplus}
\ee
providing a $G$-invariant, integrable K\"ahler structure to the orbit.

Moreover, to find which of the exceptional coadjoint orbits allow the existence of a K\"ahler structure, we recall that we must require $n=1$ in \eqref{eq:n1-exceptional-b0-2}, since for $n>1$ the obstruction for the existence of a K\"ahler structure is the same as in the purely bosonic setting.
The case $n=1$ is the vacuum orbit with geometry
\be
\mathcal{O}_{b_0=-\frac{c}{48\pi}}\cong \mathrm{Diff}_{\mathrm{sc}}\,\mathrm{S}^{1|1}/(\OSp(1|2)/\ZZ_2)\ .
\ee
Here, the relevant supergroup is $\OSp(1|2)/\ZZ_2$, whose bosonic subgroup is $\mathrm{SL}(2,\RR)$. Its Lie algebra is $\mathfrak{osp}(1|2)=\mathrm{span}\{L_0, L_{\pm 1}, G_{\pm \frac{1}{2}}\}$. The K\"ahler structure of the vacuum orbit reads
\be
\mathcal{K}_{\vac}^+=\mathrm{span}\left\{L_{n}, G_{r}
: n\geq 2, \; r \geq \frac{3}{2}\right\}\ . \label{eq:n1-virasoro-vacuum-kplus}
\ee
 In the Ramond-sector, the case $b_0=0$ also includes the problematic mode $G_0$ in the stabilizer, and we can define the orbit
\be
\mathcal{O}_{b_0=0}\cong \mathrm{Diff}_{\mathrm{sc}}\,\mathrm{S}^{1|1}/\mathrm{S}^{1|1}\ ,
\ee
which does carry a K\"ahler structure, given by
\be
\mathcal{K}_{\mR}^+=\mathrm{span}\{L_{n} ,\, G_{r}: n\geq 1,  \; r \ge 1\}\ . \label{eq:n1-virasoro-ramond-kplus}
\ee
For any other value of $b_0$, the Ramond-sector does not admit an orbit with an appropriate K\"ahler structure because of the $G_0$ mode. 
\paragraph{Evaluation of the superindex theorem.}
With these preparations, we can apply the superindex theorem to the prequantum line bundle $\mathcal{L}$ defined over the coadjoint orbits described previously. By the same logic of equivariant localization, such an index can always be reduced to local contributions from fixed points on the \emph{bosonic} submanifold, see Appendix~\ref{app:reducing-superindex-theorem}. The character formula \eqref{eq:super-character-formula} gives
\be
 \chi_\pm (t)=\operatorname{(s)tr}(q^{L_0 -\frac{c}{24}})=\sum_{p \text{ fixed}}\e^{\mu_p(X)}\frac{\prod_{j }(1\pm\e^{-\xi_jt })}{\prod_{i }(1-\e^{-x_it })}\ , \label{eq:n1-superindex}
\ee
where the $x_i$ and $\xi_j$ are eigenvalues associated with the bosonic and fermionic parts of the infinitesimal action on the holomorphic tangent spaces of the fixed points. We also wrote $q=\e^{t}$ as in the bosonic case with $t$ the equivariant parameter. In the supersymmetric setting, we can compute the supercharacter (with $(-1)^\text{F}$ insertion) or the ordinary character (without $(-1)^\text{F}$ insertion), corresponding to the $\pm$ sign in the character.

To compute the eigenvalues $x_i$ and $\xi_j$, let us note that from \eqref{eq:n1-coadjoint-orbit-action}, the action of $a \in \U(1)$ generated by $L_0$ on a coadjoint vector is given by 
\be
    t \cdot (\beta(\varphi)+\theta b(\varphi))=\beta(\varphi+t)+\theta b(\varphi+t)\ .
\ee
Here $\beta$ is a $\frac{3}{2}$-differential and $b$ a quadratic differential. The spectrum of $L_0$ is given by $x_i\in \ZZ$ and $\xi_j \in  \ZZ+\delta$. The eigenvalues that contribute to \eqref{eq:n1-superindex} are those associated to the holomorphic directions $\mathcal{K}^+$, corresponding to the three cases \eqref{eq:n1-virasoro-generic-kplus}, \eqref{eq:n1-virasoro-vacuum-kplus} and \eqref{eq:n1-virasoro-ramond-kplus}.
 
The $\U(1)$ action admits a unique fixed point in each orbit given by the constant coadjoint vector $B=\theta b_0$. Analogously to \eqref{eq:momentum-map-virasoro}, the momentum map is given by 
\be
\mu_{B}(t)=t\int_{\mathrm{S}^1} \d\varphi\, \d\theta\; \theta b_0=2\pi t b_0\ ,
\ee
where again, through the pairing with the $L_0$ generator, we can identify $2\pi b_0=h-\frac{c}{24}$.
 
It follows that the character formulas in the three cases of the generic NS-sector orbit, the vacuum orbit and the Ramond ground state orbit are
\begin{subequations}
\begin{align}
\chi_{\pm,h}(t)&=q^{h-\frac{c}{24}}\prod_{m=1}^\infty \frac{ 1\pm q^{m-\frac{1}{2}}}{1-q^m}\ , \\
    \chi_{\pm,\vac}(t)&=q^{-\frac{c}{24}}\prod_{m=2}^\infty \frac{ 1\pm q^{m-\frac{1}{2}}}{1-q^m}\ , \\
    \chi_{\pm,\mR}(t)&=\prod_{m=1}^\infty \frac{ 1\pm q^m}{1-q^m}\ .
\end{align} \label{eq:N1-Virasoro-characters}%
\end{subequations}
Notice in particular that the conformal weight of the Ramond ground state correctly becomes $h=\frac{c}{24}$ and that for the supercharacter, it becomes $\chi_{-,\mR}(t)=1$, which is the Witten index of the representation.

 \subsubsection{\texorpdfstring{$\mathcal{N}=2$}{N=2} Super Virasoro algebra}
\paragraph{Definition.}
The $\mathcal{N}=2$ super Virasoro algebra extends the $\mathcal{N}=1$ case by including a $\U(1)$ R-symmetry current and a corresponding superpartner. Hence, the $\mathcal{N}=2$ Virasoro algebra $\mathfrak{vir}_{\mathcal{N}=2}$ is generated by the following modes:
\begin{itemize}
    \item The Virasoro generators $L_n$, with $n \in \ZZ$,
    \item The $\u(1)$ current algebra generators $J_n$, with $n \in \ZZ$,
    \item Two supercurrents $G_r^+$ and $G_r^-$, where $r \in \ZZ + 1/2$ in the Neveu--Schwarz (NS) sector, or $r \in \ZZ$ in the Ramond (R) sector.
\end{itemize}
Note that we can in principle consider a compact or non-compact R-symmetry. We will consider the non-compact version which does not quantize the corresponding charges. From the point of view of the Virasoro group, the non-compact version leads to a connected group, while for compact radius, we also get winding sectors which are disconnected from the identity group element.

There is an additional feature called spectral flow \cite{Schwimmer_Seiberg}. Spectral flow is an outer Lie algebra automorphism that relates different sectors of the theory, particularly the Neveu--Schwarz and Ramond sectors. In the compact version of the $\U(1)$ R-symmetry, it can be realized by conjugation with a particular winding element, but the automorphism is genuinely outer in the non-compact case. Spectral flow acts non-trivially on the set of representations. In the following discussions of the $\mathcal{N}=2, 4$ Virasoro algebra, we will focus only on the NS-sector since, through spectral flow, one can obtain the corresponding statements for the R-sector.

The geometric definition of the $\mathcal{N}=2$ Virasoro algebra is, as one might expect, related to the group $\widehat{\mathrm{Diff}_{\mathrm{sc}}\,\mathrm{S}^{1|2}}$ of centrally extended superconformal diffeomorphisms of the $\mathcal{N}=2$ supercircle $\mathrm{S}^{1|2}$, defined below. This supercircle carries the bosonic angular coordinate $\varphi$ and two additional fermionic ones, $\theta^+$ and $\theta^-$, and so a general superdiffeomorphism $(\Phi, \Theta^+, \Theta^-)$ is given by three functions that encode the reparametrization of each coordinate.
\paragraph{$\mathcal{N}=2$ Superconformal diffeomorphisms.}
As before, we want to restrict our discussion to superconformal diffeomorphisms, which are the ones preserving the $(0|2)$ distribution generated by the two super derivatives,
\be
D_\pm = \partial_{\theta^\mp} + \theta^\pm \partial_\varphi\ .
\ee
This is equivalent to demanding that the superdiffeomorphisms are also contactomorphisms with respect to the contact form $\alpha_{\mathcal{N}=2}=\d\varphi-\theta^+\d\theta^--\theta^-\d\theta^+$.
Analogously to \eqref{eq:n1-superconformal-vector-field}, the contact condition relates the three components of a general
vector field on $\mathrm{S}^{1|2}$. The resulting superconformal vector fields are parametrized by a
single generating superfunction $F$.
\be  
F(\varphi,\theta^+,\theta^-)=f(\varphi)+\theta^+ f_+(\varphi)+\theta^-f_-(\varphi)+\theta^+ \theta^-f_{+-}(\varphi)\ .
\ee
We can then Fourier expand the different components, with the modes of $f_{+-}(\varphi)$ being the modes of the R-symmetry current, the modes of $f_\pm(\varphi)$ the modes $G_r^\pm$ and the modes of $f(\varphi)$ the Virasoro modes.
In this way, the Lie algebra of centrally extended superconformal vector fields provides a realization of $\mathfrak{vir}_{\mathcal{N}=2}$ when endowed with a commutator that generalizes the definition given in equation~\eqref{eq:n1-supercommutator}, as given in \cite{Delius:1990pt}. 
\paragraph{Coadjoint orbits.}
By studying the infinitesimal coadjoint action on a constant dual element $B_0=A_0+\theta^+ \theta^- b_0$, we can find the possible stabilizers associated with coadjoint orbit geometries containing $B_0$, as carried out in \cite{Yang:1991rd}. Here, $A_0$ should be viewed as a constant $\U(1)$ gauge field around the circle. Applying the criterion \eqref{eq:witten-criterion} for the existence of a K\"ahler structure, we arrive at the following possibilities. For generic values of $A_0$ and $b_0$, the orbits are 
\be
\mathcal{O}_{B_0}\cong \mathrm{Diff}_{\mathrm{sc}}\,\mathrm{S}^{1|2}/(\mathrm{S}^1\times \RR)\ ,
\ee
where the stabilizer is generated by $L_0$ and $J_0$.

For $B_0= -\frac{c}{48\pi}\theta^+ \theta^-$ in the NS-sector, we find the vacuum orbit 
\be
\mathcal{O}_{B_0} \cong \mathrm{Diff}_{\mathrm{sc}}\,\mathrm{S}^{1|2}/\widetilde{(\OSp(2|2)/\ZZ_2)}\ ,
\ee
where the Lie algebra $\mathfrak{osp}(2|2)$ is generated by the modes $\{J_0, L_0, L_{\pm1}, G_{\pm 1/2}^+, G_{\pm 1/2}^-\}$. We need to consider the universal cover to correctly account for the non-compactness of the R-symmetry. Finally, it is possible to tune the values of $A_0$ and $b_0$ in such a way that the orbit stabilizers have more generators than the generic orbits but fewer than the vacuum. In fact, for 
\be\label{eq:n2-bps-location}
b_0= \pm \frac{1}{2}A_0- \frac{c}{48\pi}\ ,
\ee
these BPS (chiral primary) and anti-BPS (antichiral primary) orbits are given by 
\be\label{eq:n2-bps-orbits}
\mathcal{O}_{B_0}\cong \mathrm{Diff}_{\mathrm{sc}}\,\mathrm{S}^{1|2}/\U(1|1)^{\pm}\ .
\ee
Here $\U(1|1)^+$ is generated by the modes $\{L_0, J_0, G_{-1/2}^+, G_{1/2}^-\}$ and $\U(1|1)^-$ is generated by the modes $\{L_0, J_0, G_{1/2}^+, G_{-1/2}^-\}$.
Quantization of the orbit
through $B_0$ gives a highest-weight representation with highest-weight state
$\ket{h,Q}$, where
\be
L_0\ket{h,Q}=h\ket{h,Q}\ ,
\qquad
J_0\ket{h,Q}=Q\ket{h,Q}\ .
\ee
Using the pairing between $\mathfrak{vir}_{\mathcal{N}=2}$ and its dual, these quantities are related to $b_0$ and $A_0$ by
\be\label{eq:n2-hq-identification}
2\pi b_0=h-\frac{c}{24}\ ,
\qquad
2\pi A_0=Q\ .
\ee
Using this identification, we uncover the BPS nature of the orbits \eqref{eq:n2-bps-orbits}, since \eqref{eq:n2-bps-location} implies the saturation of the (anti-)BPS bounds $h \geq \pm \frac{1}{2}Q$. 
\paragraph{K\"ahler structures.}
For all of the accounted geometries, the holomorphic directions are spanned by the annihilation (positive) modes not contained in the coadjoint orbit stabilizer, i.e.
\begin{subequations}
\begin{align}
\mathcal K_{\gen}^+
&= \operatorname{span}\{L_{n},J_{n},G_{r}^+,G_{r}^-:\ n\geq1,\ r\geq\tfrac{1}{2}\}, \\[3pt]
\mathcal K_{\vac}^+
&= \operatorname{span}\{L_{n},J_{1},J_{n},G_{r}^+,G_{r}^-:\ n\geq2,\ r\geq\tfrac{3}{2}\}, \\[3pt]
\mathcal K_{\BPS}^+
&= \operatorname{span}\{L_{n},J_{n},G_{1/2}^+,G_{r}^+,G_{r}^-:\ n\geq1,\ r\geq\tfrac{3}{2}\}, \\[3pt]
\mathcal K_{\text{anti-}\BPS}^+
&= \operatorname{span}\{L_{n},J_{n},G_{1/2}^-,G_{r}^+,G_{r}^-:\ n\geq1,\ r\geq\tfrac{3}{2}\}\ .
\end{align}
\end{subequations}
Shortly, these holomorphic directions will be used to select the eigenvalues of the infinitesimal toric action that will contribute to the localized superindex theorem.

\paragraph{Fixed points.}
In this setting, the toric action used for localization is generated by $L_0$ and $J_0$; for $(t,s)\in \U(1) \times \RR$ it is explicitly given by
\begin{multline}
    (t,s) \cdot (A(\varphi)+\theta^+ \chi_+(\varphi)+\theta^-\chi_-(\varphi)+\theta^+ \theta^- b(\varphi))\\
    =A(\varphi+t)+\e^{-is}\theta^+ \chi_+(\varphi+t)+\e^{is}\theta^-\chi_-(\varphi+t)+\theta^+ \theta^-b(\varphi+t)\ , \label{eq:n2-toric-action}
\end{multline}
where $A,\chi_\pm, b$ define the coadjoint vector in $\mathfrak{vir}_{\mathcal{N}=2}^*$. Notice that $A$ transforms as a gauge field, $\chi_\pm$ are $\frac{3}{2}$-differentials and $b$ a quadratic differential. As before, such an action provides a unique fixed point in each coadjoint orbit of the form $B_0=A_0+\theta^+ \theta^-b_0$, with $A_0$ and $b_0$ constant.
\paragraph{Evaluation of the superindex theorem.} We now proceed as before, using the equivariant superindex theorem as described in Appendix~\ref{app:reducing-superindex-theorem} to compute the character of the representation associated with each orbit. As in the $\mathcal{N}=1$ case \eqref{eq:n1-superindex}, the character is obtained as a sum over the fixed points of the toric action,
\be
\chi_\pm(t,s)=\operatorname{(s)tr}\big(q^{L_0-\frac{c}{24}}y^{J_0}\big)=\sum_{p\text{ fixed}}\e^{\mu_p}\,\frac{\prod_j\big(1\pm \e^{-\xi_j}\big)}{\prod_i\big(1-\e^{-x_i}\big)}\ , \label{eq:n2-superindex}
\ee
where $x_i$ and $\xi_j$ are the eigenvalues of the infinitesimal $(L_0,J_0)$ action on the holomorphic directions $\mathcal{K}^+$ and $\mu_p$ is the value of the momentum map at the fixed point, all determined below. We defined $q=\e^{t}$ and $y=\e^{s}$.

For $(t,s) \in \u(1) \times \RR$ and $B=A+\theta^+\chi_+ + \theta^-\chi_- + \theta^+ \theta^-b$ an element of $\mathfrak{vir}_{\mathcal{N}=2}^*$, the infinitesimal version of \eqref{eq:n2-toric-action} reads
\begin{align}
        \delta_{(t,s)}B=it\partial_\varphi A+\theta^+ i(t\partial_\varphi + is)\chi_++\theta^-i(t\partial_\varphi - is)\chi_-+\theta^+\theta^-it\partial_\varphi b\ ,
\end{align}
whose spectrum for $(L_0,J_0)$ in the NS-sector is given by
\be 
\{(n,0)^2\}_{n \le -1} \big|\{(r,\pm 1)\}_{r \le -1/2}\ ,
\ee
where we separated the bosonic and fermionic eigenvalues by a slash and denote multiplicity by a superscript. The bosonic eigenvalues originate from both $A$ and $b$ and the fermionic ones from $\chi_+$ and $\chi_-$.

Before constructing the character formulas, we describe the momentum map associated with the Hamiltonian $\U(1) \times \RR$ action. At a point $B(\varphi)=A(\varphi)+\theta^+ \theta^-b(\varphi)$ in the coadjoint orbit, the momentum map contracted with $(t,s)\in \u(1)\times \RR$ is 
\begin{equation}
    \mu_B(t,s)=s\int_{\mathrm{S}^1}\d\varphi\, A(\varphi)+t\int_{\mathrm{S}^1}\d\varphi\, b(\varphi)\ .
\end{equation}
For points $B_0=A_0+\theta^+\theta^-b_0$ constant in $\varphi$, the value of the momentum map is $\mu_{B_0}(t,s)=2\pi t b_0+2\pi s A_0$. For generic orbits, we can then use the $\mathcal{K}^+$ holomorphic modes together with the identification \eqref{eq:n2-hq-identification} to write
\be
\chi_{\pm,h,Q}(t,s)=q^{h-\frac{c}{24}}y^{Q}\prod_{m=1}^{\infty}\frac{(1\pm q^{m-\frac{1}{2}}y)(1\pm q^{m-\frac{1}{2}}y^{-1})}{(1-q^{m})^2}\ . \label{eq:N2-Virasoro-generic}
\ee
Analogously, this procedure can be used to compute the character formulas for the vacuum representation, as well as the (anti-)BPS representations:
\begin{subequations}
\begin{align}
    \chi_{\pm,\vac}(t,s)&=q^{-\frac{c}{24}}\prod_{m=2}^{\infty}\frac{(1\pm q^{m-\frac{1}{2}}y)(1\pm q^{m-\frac{1}{2}}y^{-1})}{(1-q^{m})(1-q^{m-1})}\ , \\
     \chi_{\pm,\BPS,h}(t,s)&=q^{h-\frac{c}{24}}y^{2h}\prod_{m=1}^\infty \frac{(1\pm q^{m-\frac{1}{2}}y)(1\pm q^{m+\frac{1}{2}}y^{-1})}{(1-q^m)^2}\ , \\
     \chi_{\pm,\text{anti-}\BPS,h}(t,s)&=q^{h-\frac{c}{24}}y^{-2h}\prod_{m=1}^\infty \frac{(1\pm q^{m+\frac{1}{2}}y)(1\pm q^{m-\frac{1}{2}}y^{-1})}{(1-q^m)^2}\ .
\end{align} \label{eq:N2-Virasoro-special}%
\end{subequations}
All the character formulas above are in complete agreement with the expressions previously found by Eguchi and Taormina in \cite{Eguchi:1988af}.

\subsubsection{\texorpdfstring{$\mathcal{N}=4$}{N=4} Super Virasoro algebra}

\paragraph{Definition.}
There are two versions of the $\mathcal{N}=4$ superconformal algebra: \emph{small} and \emph{large}. We focus on the small $\mathcal{N}=4$ algebra, which includes:
\begin{itemize}
    \item The Virasoro generators $L_n$,
    \item An $\SU(2)$ R-symmetry current algebra generated by $J_n^a$, $a\in\{3,\pm\}$,
    \item Four supercurrents $G_r^{\alpha \dot{\alpha}}$ with $\alpha \in \{\pm\}$ the $\SU(2)$ R-symmetry spinor index and $\dot{\alpha} \in \{\pm\}$ the outer automorphism spinor index, with $r \in \ZZ$ in the R-sector or $r \in \ZZ + \frac{1}{2}$ in the NS-sector.
\end{itemize}
The level $k$ of the $\su(2)$ R-symmetry current is related to the central charge as $c=6k$.
This super-extension of the Virasoro algebra can be realized geometrically by considering the central extension of the superconformal diffeomorphism group of the supercircle $\mathrm{S}^{1|4}$ with coordinates $(\varphi\,|\,\theta^{\alpha\dot{\alpha}})$.
\paragraph{$\mathcal{N}=4$ Superconformal diffeomorphisms.}
As a natural generalization of superconformal diffeomorphisms in the $\mathcal{N}=1,2$ setting, $\mathcal{N}=4$ superconformal diffeomorphisms are superdiffeomorphisms $(\Phi, \Theta^{\alpha \dot \alpha})$ that preserve the $(0|4)$ distribution generated by all super derivatives
\be
D_{\alpha \dot \alpha}=\partial_{\theta^{\alpha \dot \alpha}}+\epsilon_{\alpha\beta}\epsilon_{\dot\alpha\dot\beta}\theta^{\beta \dot \beta}\partial_\varphi\ .
\ee
Such diffeomorphisms are the ones preserving the expected contact structure \be
\alpha_{\mathcal{N}=4}=\d\varphi-\epsilon_{\alpha\beta}\epsilon_{\dot\alpha\dot\beta}\theta^{\alpha \dot \alpha}\d\theta^{\beta \dot \beta}.\ee With an appropriate Lie bracket we have $\mathfrak{vir}_{\mathcal{N}=4}=\mathrm{Lie}(\widehat{\mathrm{Diff}_{\mathrm{sc}}\,\mathrm{S}^{1|4}})$.
As before, the superconformal condition can be used to write any superconformal vector field in terms of a single superfield $f(\varphi,\theta^{\alpha\dot\alpha})$ which additionally satisfies the shortening constraint \cite{Matsuda:1989kp}
\be
(\sigma^{\dot a})^{\dot\alpha\dot\beta}\,D_{\alpha\dot\alpha}D_{\beta\dot\beta} f=0\ ,\qquad \dot a \in \{3,\pm\}\ , \label{eq:shortening-constraint}
\ee
with $\sigma^{\dot a}$ the Pauli matrices acting on the outer (dotted) indices; equivalently, the part of $D_{\alpha\dot\alpha}D_{\beta\dot\beta}f$ symmetric in the outer indices $\dot\alpha\dot\beta$ vanishes. Its component expansion reads
\begin{multline}
         f(\varphi, \boldsymbol{\theta})=f_0(\varphi)+f_{\alpha\dot\alpha}(\varphi)\,\theta^{\alpha\dot\alpha}+f_{\alpha\dot\alpha;\beta\dot\beta}(\varphi)\,\theta^{\alpha\dot\alpha}\theta^{\beta\dot\beta}+f_{\alpha\dot\alpha;\beta\dot\beta;\gamma\dot\gamma}(\varphi)\,\theta^{\alpha\dot\alpha}\theta^{\beta\dot\beta}\theta^{\gamma\dot\gamma}\\
         +f_{\alpha\dot\alpha;\beta\dot\beta;\gamma\dot\gamma;\delta\dot{\delta}}(\varphi)\,\theta^{\alpha\dot\alpha}\theta^{\beta\dot\beta}\theta^{\gamma\dot\gamma}\theta^{\delta\dot\delta}\ ,
\end{multline}
where $\boldsymbol{\theta}$ denotes the collection $\{\theta^{\alpha\dot\alpha}\}$. Antisymmetrization in $\theta^{\alpha\dot\alpha}$ leads to $1, 4, 6, 4, 1$ components, transforming under $\SU(2)_\mR\times\SU(2)_{\out}$ (acting on the undotted and dotted indices) as
\be
\mathbf1,\quad (\mathbf2,\mathbf2),\quad (\mathbf3,\mathbf1)\oplus(\mathbf1,\mathbf3),\quad (\mathbf2,\mathbf2),\quad \mathbf1
\ee
at orders $\boldsymbol{\theta}^0,\dots,\boldsymbol{\theta}^4$. The shortening constraint \eqref{eq:shortening-constraint} removes the terms transforming in the $(\mathbf1,\mathbf3)$ at order $\boldsymbol{\theta}^2$, leaving the R-symmetry triplet $(\mathbf3,\mathbf1)$ as the only spin-one fields, while at higher orders it fixes the $\boldsymbol{\theta}^3$ and $\boldsymbol{\theta}^4$ coefficients to be $\partial_\varphi$-descendants of the $\boldsymbol{\theta}^1$ and $\boldsymbol{\theta}^0$ ones. The independent content is therefore $\mathbf1\oplus(\mathbf2,\mathbf2)\oplus(\mathbf3,\mathbf1)$, whose Fourier coefficients give the modes $L_n$, $G_r^{\alpha\dot\alpha}$ and $J_n^a$ of the small $\mathcal{N}=4$ superconformal algebra.
\paragraph{Coadjoint orbits.} 
Dually, a general coadjoint vector $B \in \mathfrak{vir}_{\mathcal{N}=4}^*$ can also be expanded in components. The corresponding dual shortening conditions leave an $\SU(2)_\mR$ gauge field $A^a$ at order $\boldsymbol{\theta}^2$, the gravitino background $\chi^{\alpha\dot \alpha}$ at order $\boldsymbol{\theta}^3$, and the stress tensor (quadratic differential) at order $\boldsymbol{\theta}^4$, while removing the weight-$\tfrac{1}{2}$ fermions ($\boldsymbol{\theta}^1$) and the weight-$0$ scalar ($\boldsymbol{\theta}^0$).
The coadjoint orbits passing through a constant Grassmann-even element $B_0 \in \mathfrak{vir}_{\mathcal{N}=4}^*$ are of the type $\mathrm{Diff}_{\mathrm{sc}}\,\mathrm{S}^{1|4}/H$ for some stabilizer $H$, as previously studied in \cite{Aoyama:2018lfc, Aoyama:2018voj}. Generic values of $B_0$ provide the stabilizer $H=\mathrm{S}^1 \times \mathrm{S}^1$ generated by $L_0$ and $J_0^3$. The stabilizer of the spin-0 representation is $H=\mathrm{S}^1 \times \SU(2)$ generated by $L_0$ and $J_0^a$. The stabilizer associated with the vacuum orbit is $H=\PSU(1,1|2)$, whose Lie algebra generators are $\{L_0, L_{\pm1}, J_0^\pm, J_0^3, G_{\pm1/2}^{\pm \pm}\}$. The last orbit corresponds to the BPS orbit. Its stabilizer is $H=(\SU(1|1)\times \SU(1|1))^+$, generated by $\{L_0, J_0^3, G_{-1/2}^{+\dot\alpha}, G_{+1/2}^{-\dot\alpha}\}$ for $\dot\alpha\in\{\pm\}$. Notice that the subgroup generated by $\{L_0, J_0^3, G_{+1/2}^{+\dot\alpha}, G_{-1/2}^{-\dot\alpha}\}$ for $\dot\alpha\in\{\pm\}$ is related by a Weyl reflection in $\SU(2)$, which is part of the inner automorphism group of $\PSU(1,1|2)$ and thus leads to an equivalent orbit.
\paragraph{K\"ahler structures.} 
Each orbit described above can be endowed with an appropriate K\"ahler structure which is spanned by the annihilation modes not contained in the stabilizer,
\begin{subequations}\label{eq:n4-kahler-structures}
\begin{align}
\mathcal K_{\gen}^+
&=
\operatorname{span}
\{L_{n},J_0^+,J_{n}^a,G_{r}^{\alpha\dot\alpha}:
n\geq1,\ r\geq\tfrac{1}{2},\ a \in \{3,\pm \},\ \alpha,\dot\alpha\in\{\pm\}\}\ ,
\\
\mathcal K_{0}^+
&=
\operatorname{span}
\{L_{n},J_{n}^a,G_{r}^{\alpha\dot\alpha}:
n\geq1,\ r\geq\tfrac{1}{2},\ a \in \{3,\pm \},\ \alpha,\dot\alpha\in\{\pm\}\}\ ,
\\
\mathcal K_{\vac}^+
&=
\operatorname{span}
\{L_{n},J_{1}^a,J_{n}^a,G_{r}^{\alpha\dot\alpha}:
n\geq2,\ r\geq\tfrac{3}{2},\ a \in \{3,\pm\},\ \alpha,\dot\alpha\in \{\pm\}\}\ ,
\\
\mathcal K_{\BPS}^+
&=
\operatorname{span}
\{L_{n},J_0^+,J_{n}^a,
G_{1/2}^{+\dot\alpha}, G_{r}^{\alpha\dot\alpha}:
n\geq1,\ r\geq\tfrac{3}{2},\ a \in \{3,\pm\},\ \alpha,\dot\alpha\in \{\pm\}\}\ .
\end{align}
\end{subequations}
 \paragraph{Fixed points.}
In this setting, there are three commuting $\U(1)$ actions on the coadjoint orbits, which will be used to localize the integral form of the character formula. We describe each of these actions by specifying how they act on each coordinate.
  \begin{itemize}
      \item The first $\U(1)$ action is the usual one, generated by $L_0$,
      \begin{equation}
         f(\varphi, \theta^{\alpha\dot\alpha})\mapsto f(\varphi+t, \theta^{\alpha\dot\alpha})\ .
      \end{equation}
      \item The second $\U(1)$ action is generated by $J_0^3$ and acts trivially on the $\varphi$ coordinate. It acts on the bosonic coordinates by conjugation of the gauge field as in \eqref{eq:loop-group-toric-action} and on the fermionic coordinates according to their $\SU(2)$ R-symmetry index $\alpha$: for $s \in \mathfrak{u}(1)$,
      \begin{align}
              &\theta^{+\dot\alpha} \mapsto \e^{s/2}\,\theta^{+\dot\alpha}\ , \qquad 
              \theta^{-\dot\alpha} \mapsto \e^{-s/2}\,\theta^{-\dot\alpha}
      \end{align}
      for both values of $\dot\alpha\in\{\pm\}$.
      \item There is an additional $\U(1)$ charge associated with the outer automorphism of the supersymmetry generators. It acts trivially on the bosonic coordinate and acts on the fermionic coordinates according to their outer automorphism index $\dot\alpha$: for $u \in \mathfrak{u}(1)$,
     \begin{align}
              \theta^{\alpha \dot+} \mapsto \e^{u/2}\,\theta^{\alpha \dot+}\ , \qquad 
              \theta^{\alpha \dot-} \mapsto \e^{-u/2}\,\theta^{\alpha \dot-}
      \end{align}
      for both values of $\alpha\in\{\pm\}$.
  \end{itemize}
For the supersymmetric localization, we have to determine the fixed points of the $\U(1)^3$ action on the bosonic subspace of the coadjoint orbit, which consists of the action on quadratic differentials as discussed in \eqref{eq:virasoro-coadjoint-action} and an $\SU(2)$ gauge field as in \eqref{eq:loop-group-toric-action}. We already classified all the fixed points there. The $\U(1)$ action generated by $L_0$ forces the gauge field and the quadratic differential to be constant, while the $\U(1)$ action generated by $J_0^3$ forces the gauge field to take the form as in \eqref{eq:fixed-points-kac-moody}. 
Finally, the outer $\U(1)_\out$ acts only on the fermionic directions. Since all fermionic tangent directions have non-zero equivariant weight under the torus action, they contribute to the Euler and Todd class but do not alter the fixed-point set.
The fixed points are thus in one-to-one correspondence with the affine $\SU(2)$ fixed points. In the Kac--Moody case, we only discussed the generic character case. For the spin-0 orbit and the vacuum orbit, the two sign choices in \eqref{eq:fixed-points-kac-moody} correspond to the same set of fixed points and thus fixed points are only indexed by an integer.

\paragraph{Charges on the tangent bundle.}
For the superindex theorem \eqref{eq:super-character-formula}, we have to evaluate the eigenvalues of the $\U(1)^3$ action on the holomorphic tangent space. This amounts to noting the eigenvalues of the modes appearing in \eqref{eq:n4-kahler-structures} for the generators $(L_0,J_0^3,Q_\out)$, with $Q_\out$ the charge with respect to the outer $\U(1)$ symmetry. They take the form
\begin{subequations}
    \begin{align}
        \gen&: \ \{(n,0,0)^2\}_{n\le -1} \cup \{(n,1,0)\}_{n\le 0} \cup \{(n,-1,0)\}_{n\le -1}  \big| \{(r,\pm \tfrac{1}{2},\pm  \tfrac{1}{2})\}_{r \le -1/2}\ , \\
        0&: \ \{(n,0,0)^2\}_{n\le -1} \cup \{(n,\pm 1,0)\}_{n\le -1}  \big| \{(r,\pm \tfrac{1}{2},\pm  \tfrac{1}{2})\}_{r \le -1/2}\ , \\
        \vac&: \ \{(-1,0,0)\} \cup \{(n,0,0)^2\}_{n\le -2} \cup \{(n,\pm 1,0)\}_{n\le -1}  \big| \{(r,\pm \tfrac{1}{2},\pm  \tfrac{1}{2})\}_{r \le -3/2}\ , \\
        \BPS&: \ \{(n,0,0)^2\}_{n\le -1} \cup \{(n,1,0)\}_{n\le 0} \cup \{(n,-1,0)\}_{n\le -1} \big| \{(-\tfrac{1}{2},\tfrac{1}{2},\pm \tfrac{1}{2} )\} \nonumber\\
        &\qquad\qquad\qquad\qquad\cup \{(r,\pm \tfrac{1}{2},\pm  \tfrac{1}{2})\}_{r \le -3/2}\ ,
    \end{align} \label{eq:n4-charges}%
\end{subequations}
where we separated the eigenvalues on the bosonic directions from those on the fermionic directions with a slash. Superscripts denote multiplicities. This assignment of charges holds at the fixed point that we chose to label the coadjoint orbit, but not necessarily at the other fixed points. For the generic orbit, the stabilizer $H=\mathrm{S}^1 \times \mathrm{S}^1$ of the orbit is invariant under the action of the affine Weyl group which means that in this case, this set of charges is correct for all fixed points. The discussion in that case is in parallel to Section~\ref{subsec:affine-kac-moody}. For the other two orbits, some charges are missing. To figure out the charges at the other fixed points, we will need to remove these missing charges from the fixed points. These missing charges are not invariant under the affine Weyl group. Instead, we can obtain them by noticing that the Weyl group is generated by the $\SU(2)$ Weyl group and spectral flow, which maps $J_0^\pm \to J_{\pm 2n}^\pm$ and $G_r^{\pm\dot \alpha} \to G_{r \pm n}^{\pm\dot\alpha}$. 

\paragraph{Evaluation of the superindex theorem in the generic case.} To compute the character formula for the $\mathcal{N}=4$ super Virasoro algebra \cite{Eguchi:1987wf, Eguchi:1988af} we will proceed as we did for the Weyl--Kac formula. It is convenient to use the super $\widehat{A}$-class with the appropriate correction of the Chern character of the square root of the canonical line bundle and the corresponding correction from the fermions. The reason we need to apply this procedure is exactly the same as before: the toric action admits infinitely many fixed points, and replacing the super Todd class with the super $\widehat{A}$ ensures that all fixed points give essentially the same contribution for the Todd class, the Chern characters and the Euler classes. For the generic orbit, this statement is correct up to a sign, because the stabilizer $H=\mathrm{S}^1 \times \mathrm{S}^1$ of the orbit is invariant under the action of the affine Weyl group and is completely in parallel to the discussion in Section~\ref{subsec:affine-kac-moody}. For the others, we will need to remove the missing charges in \eqref{eq:n4-charges} by hand from the tangent spaces around the fixed points and thus we will continue to discuss first the generic character.

The correction when passing to the $\widehat{A}$ class shifts $j \mapsto j+\frac{1}{2}$ and $k \mapsto k+1$. This shift in $j$ is induced by the Weyl vector as described in Section~\ref{subsec:affine-kac-moody} and Appendix~\ref{subapp:affine-ghosts}, while the shift in $k$ is different from the case of $\widehat{\su(2)}_k$ where the correction yielded $k \mapsto k+2$. This modification can again be computed from the computation of the canonical line bundle, see Appendix~\ref{subapp:n4-ghosts}.
Then, the localized index theorem \eqref{eq:super-character-formula} yields for the generic character
\begin{align}\label{eq:n4-localized-index}
        \chi_{\pm,h,j,k}(t,s,u)
=\left(\e^{\frac{1}{2}\sum_j \xi_j-\frac{1}{2}\sum_i x_i}\frac{\prod_j (1\pm\e^{-\xi_j})}{\prod_{i}(1-\e^{-x_i})}\right)\sum_{p \text{ fixed}} (\epsilon_p \, \e^{\mu_p})\ ,
\end{align}
where $\mu_p$ is the momentum map with shifted $j$ and $k$ and $\epsilon_p$ is the sign associated to the fixed points, which is the same as in the Kac--Moody case \eqref{eq:su2-sum-over-fixed-points}.
Noting that the bosonic modes of the $\mathcal{N}=4$ Virasoro algebra reproduce copies of $\widehat{\su(2)}_k$ and $\mathfrak{vir}$, we can relate the momentum maps,
\begin{equation}
    \mu_{\mathcal{N}=4} = \mu_{\mathfrak{vir}}+\mu_{\widehat{\su(2)}_k}\ , \label{eq:N4-momentum-map}
\end{equation}
where $\mu_{\mathfrak{vir}}$ corresponds to the momentum map \eqref{eq:momentum-map-virasoro} defined for the Virasoro orbits and $\mu_{\widehat{\su(2)}_k}$ is the momentum map \eqref{eq:momentum-map-kac-moody} associated with the $\widehat{L\SU(2)}$ orbits. This means that \eqref{eq:su2-sum-over-fixed-points} holds essentially unchanged, except that $\e^{\mu_\mathfrak{vir}}$ has an additional contribution $q^{\delta h-\frac{k-1}{4}}$, with $\delta h=h-\frac{(j+\frac{1}{2})^2}{k+1}$ the difference of the actual conformal weight and the conformal weight provided by $\mu_{\widehat{\mathfrak{su}(2)}_k}$ in \eqref{eq:N4-momentum-map}. The subtraction $-\frac{k-1}{4}$ accounts for the central charge $6(k-1)$ that is not already provided by the oscillator degrees of freedom.
Thus, we obtain
\begin{align}
        \sum_{p \text{ fixed}}\epsilon_p\,  \e^{\mu_p}=q^{h-\frac{(j+\frac{1}{2})^2}{k+1}-\frac{k-1}{4}} \sum_{n \in \ZZ}q^{\frac{1}{k+1}(j+\frac{1}{2}+(k+1)n)^2}\big( y^{j+\frac{1}{2}+(k+1)n}-y^{-(j+\frac{1}{2}+(k+1)n)}\big)\ ,
\end{align}
where $q=\e^{t}$ and $y=\e^{s}$ as before.

To conclude for the generic character, we compute the $\widehat{A}$ contribution by picking the eigenvalues associated with the holomorphic directions, given by \eqref{eq:n4-kahler-structures}. For the generic orbit, the infinite products in \eqref{eq:n4-localized-index} yield
\begin{align}
        \Xi_{\pm}(t,s,u)&=\e^{\frac{1}{2}\sum_{j}\xi_j-\frac{1}{2}\sum_{i}x_i}\frac{\prod_j (1\pm\e^{-\xi_j})}{\prod_i (1-\e^{-x_i})}\\
        &=\frac{\prod_{m=1}^{\infty}\prod_{\alpha,\dot\alpha=\pm}\big(1\pm q^{m-1/2}y^{\frac{1}{2}\alpha}z^{\frac{1}{2}\dot\alpha}\big)}{q^{\frac{1}{4}}(y^{\frac{1}{2}}-y^{-\frac{1}{2}})\prod_{m=1}^{\infty}(1-q^m)^2(1-yq^m)(1- y^{-1}q^{m})}\ , \label{eq:n4-xi-ns}
\end{align}
with $z=\e^{u}$. We used that the zeta-function regularized sum of eigenvalues gives $q^{-\frac{1}{4}} y^{-\frac{1}{2}}$.
The factor $y^{-\frac{1}{2}}$ makes the denominator of \eqref{eq:n4-xi-ns} antisymmetric under $y \to y^{-1}$, while $q^{-\frac{1}{4}}$ accounts for the appropriate ground state energy expected from four bosonic and four fermionic degrees of freedom. In fact, $\Xi_{\text{NS}}$ can be written in terms of theta functions and has nice modular properties. If we assemble everything and also use that $c=6k$, we obtain
\be 
\chi_{\pm,h,j,k}(t,s,u)=q^{h-\frac{k-1}{4}}\Xi_\pm(t,s,u) \sum_{n \in \ZZ}q^{(k+1)n^2+(2j+1)n}\big( y^{j+\frac{1}{2}+(k+1)n}-y^{-(j+\frac{1}{2}+(k+1)n)}\big)\ . \label{eq:n4-generic-character}
\ee
\paragraph{Evaluation of the superindex theorem in the degenerate cases.} We now discuss the degenerate cases. For those, we have to remove the superfluous modes in $\Xi_\pm(t,s,u)$. Let us begin with the BPS case. The first term in the parenthesis of \eqref{eq:n4-generic-character} for $n=0$ corresponds to the BPS state. Therefore, we have to divide by $\prod_{\dot \alpha=\pm} (1\pm q^{\frac{1}{2}}y^{\frac{1}{2}}z^{\frac{1}{2}\dot \alpha})$, which appear in \eqref{eq:n4-xi-ns}, but are excluded in \eqref{eq:n4-charges} in the BPS case. The second term in the parenthesis for $n=0$ likewise corresponds to the anti-BPS state for which we should remove the factor $\prod_{\dot \alpha=\pm} (1\pm q^{\frac{1}{2}}y^{-\frac{1}{2}}z^{\frac{1}{2}\dot \alpha})$. For the terms with general $n$, spectral flow tells us that the missing generator is $G_{n+\frac{1}{2}}^{+\dot\alpha}$ for the terms in the first parenthesis and $G_{n+\frac{1}{2}}^{-\dot\alpha}$ for the terms in the second parenthesis. We thus learn that \eqref{eq:n4-generic-character} has to be modified as follows for the BPS case:
\begin{multline}
    \chi_{\pm,j,k}^\BPS(t,s,u)=q^{j-\frac{k-1}{4}} \Xi_\pm(t,s,u) \sum_{n \in \ZZ} q^{(k+1)n^2+(2j+1)n}\\
    \times \left( \frac{y^{j+\frac{1}{2}+(k+1)n}}{\prod_{\dot \alpha=\pm} (1 \pm q^{n+\frac{1}{2}}y^{\frac{1}{2}}z^{\frac{1}{2}\dot\alpha})}-\frac{y^{-(j+\frac{1}{2}+(k+1)n)}}{\prod_{\dot \alpha=\pm} (1 \pm q^{n+\frac{1}{2}}y^{-\frac{1}{2}}z^{\frac{1}{2}\dot\alpha})}\right)\ . \label{eq:n4-bps-character}
\end{multline}
We also used that $h=j$ for BPS states.

We now move on to the spin-0 case, with stabilizer $\mathrm{S}^1 \times \SU(2)_\mR$. In this case, the second term of the parenthesis of \eqref{eq:n4-generic-character} is absent since we already noticed earlier that saddle points are only indexed by integers. For $n=0$, we have to remove the action of $J_0^+$, which is absent in the spin-0 case in \eqref{eq:n4-charges}, i.e.\ multiply by $(1-y^{-1})$. For $n \ne 0$, spectral flow tells us that we have to multiply by $(1-q^{-2n} y^{-1})$, since $J_0^-$ is negatively charged under $J_0^3$, i.e. $-2$ times the charge of the supercurrent that we encountered in the BPS case above. Thus, we find in the spin $0$-case
\be 
\chi_{\pm,h,0,k}(t,s,u)=q^{h-\frac{k-1}{4}} \Xi_\pm(t,s,u) \sum_{n \in \ZZ} q^{(k+1)n^2+n} y^{\frac{1}{2}+(k+1)n}\big(1-q^{-2n} y^{-1}\big)\ . \label{eq:n4-spin0-character}
\ee
This actually agrees with \eqref{eq:n4-generic-character} for $j=0$. To see this, one has to expand the parenthesis and rename $n \to -n$ in the second term. In view of the discussion at the end of Section~\ref{subsec:affine-kac-moody}, this is not surprising. Since the enhanced stabilizer group is compact, we could have worked with the generic orbit and a degenerated symplectic form to deduce the same result.

Finally, we treat the vacuum representation, for which null-vectors are again only labelled by a single integer. For $n=0$, we have to remove the generators corresponding to the Virasoro generator $L_{1}$, supercurrents $G_{1/2}^{\alpha \dot \alpha}$ as well as $J_0^+$, which gives
\begin{align}
    \chi^\vac_{\pm,k}(t,s,u)=q^{-\frac{k-1}{4}} \Xi_\pm(t,s,u) \sum_{n \in \ZZ} \frac{q^{(k+1)n^2+n} y^{\frac{1}{2}+(k+1)n}(1-q^{-2n} y^{-1})(1-q)}{\prod_{\dot \alpha=\pm} (1 \pm q^{-n+\frac{1}{2}}y^{-\frac{1}{2}} z^{\frac{1}{2}\dot \alpha})(1 \pm q^{n+\frac{1}{2}}y^{\frac{1}{2}} z^{\frac{1}{2}\dot \alpha})}\ . \label{eq:n4-vacuum-character}
\end{align}
It is again easy to see that \eqref{eq:n4-vacuum-character} equals \eqref{eq:n4-bps-character} when specialized to $j=0$.

\acknowledgments
This paper is partially based on the Master thesis of one of us (DF) \cite{DFmasterthesis}. LE is funded by the European Union (ERC, StringScat, 101115511). Views and opinions expressed are however those of the authors only and do not necessarily reflect those of the European Union or the European Research Council Executive Agency. Neither the European Union nor the granting authority can be held responsible for them. DF is supported by FCT/Portugal and the Recovery and Resilience Plan (PRR) through projects UID/04459/2025 and UID/PRR/04459/2025.
\appendix 

\section{Reducing the superindex theorem} \label{app:reducing-superindex-theorem}
In this Appendix, we discuss the index theorem for supermanifolds. The proof of the (non-equivariant) superindex theorem is
discussed in \cite{Voronov:1990yz}, see also \cite{Eberhardt:2026hfh}.

The statement of the superindex theorem naturally generalizes the classical one. It expresses the Euler characteristic of line bundles as an integral of characteristic classes over the \emph{reduced} space obtained by setting all fermionic coordinates to zero. 
In our setting, the line bundle restricts to an ordinary line bundle over the reduced space.
Hence $c_1(\mathcal{L})$ is just an ordinary cohomology class on $\mathcal{O}_\red$, with no super-extension needed. The Todd class, however, involves both bosonic and fermionic tangent directions and is modified accordingly. Locally, the supermanifold can be modeled as the total space of a vector bundle over the reduced space. Globally, this does not hold in the complex analytic category. However, characteristic classes are insensitive to this obstruction. Thus, one can assume that the supermanifold is of this form (we say the supermanifold is split). The holomorphic tangent bundle $T \mathcal{O}$ thus splits as $T_\bos\mathcal{O}\oplus T_\ferm\mathcal{O}$. Integrating out the fermionic directions leads to the factor $\text{ch}(\bigwedge^{\!\bullet} (T_\ferm^*\mathcal{O}))$, where $\bigwedge$ represents the exterior algebra.

We can count sections of the line bundle $\mathcal{L}$ with or without insertions $(-1)^\text{F}$, meaning that we can compute $\operatorname{sdim}\, H^0(\mathcal{O},\mathcal{L})$ or $\dim\, H^0(\mathcal{O},\mathcal{L})$. We will denote the corresponding Euler characteristic by $\chi_-(\mathcal{O},\mathcal{L})$ and $\chi_+(\mathcal{O},\mathcal{L})$, respectively.  The insertion of $(-1)^\text{F}$ can be implemented on the index theorem level by considering the virtual bundle $\text{ch}(\bigwedge^{\!\bullet}_- (T_\ferm^*\mathcal{O})):=\sum_p \text{ch}((-1)^p \bigwedge^{p} (T_\ferm^*\mathcal{O}))$, since the fermionic tangent directions, and hence sections of $T_\ferm^*$, carry $(-1)^\text{F}$-charge $-1$. 

Thus, the superindex theorem in the context of coadjoint orbits of super groups reads
\be 
\chi_\pm(\mathcal{O},\mathcal{L})=\int_{\mathcal{O}_\red} \td(T_\bos \mathcal{O}) \,\text{ch}({\textstyle\bigwedge^{\!\bullet}_{\pm}} ( T_\ferm^* \mathcal{O}))\, \e^{c_1(\mathcal{L})} \ ,
\ee
with $\mathcal{O}_\red$ the reduced space, i.e.\ the bosonic submanifold of the coadjoint orbit. 
By the same reasoning, the superindex theorem can also be formulated equivariantly. 

Applying ABBV, the integral localizes to the fixed-point set of a maximal torus $T$, which for coadjoint orbits consists of isolated points. Let $x_i$ and $\xi_j$ be the equivariant Chern roots of the bosonic and fermionic tangent space, respectively, associated to holomorphic directions near a fixed point. Then by following the same manipulations as in Section~\ref{subsec:localization}, we obtain the following character formula,
\begin{align}
\chi_\pm =\sum_{p \text{ fixed}}\, \e^{\mu_p}\frac{\prod_{j}(1\pm \e^{-\xi_j})}{\prod_{i}(1-\e^{-x_i})}\ . \label{eq:super-character-formula}
\end{align}

\section{Identification of canonical line bundles} \label{app:canonical-line-bundles}
In this Appendix, we will identify the canonical line bundle of coadjoint orbits. 
Throughout, $\mathcal{O}$ denotes the coadjoint orbit equipped with the complex (or
supercomplex) structure used in the localization argument, and $\mathcal{K}_{\mathcal{O}}
:=
\det\bigl(T^{*(1,0)}\mathcal{O}\bigr)$
(or the Berezinian in the supercase)
is its canonical line bundle.

The basic input is that the functional integral of the BRST/Faddeev--Popov ghosts produces (the inverse of) the determinant line of this complex.
In the holomorphic setting this determinant line is canonically identified with
$\mathcal{K}_{\mathcal{O}}^{-1}$; this statement can be made precise using Quillen's
determinant line bundle (and its extensions to families) \cite{Quillen_linebundle, Freed_determinant}.
Equivalently, the holomorphic ghost partition function naturally takes values in
\begin{equation}
Z_{\mathrm{gh}} \in \Gamma\big(\mathcal{O},\mathcal{K}_{\mathcal{O}}^{-1}\big)\ .
\label{eq:ghost-in-k-inverse}
\end{equation}
More colloquially, the ghost system is needed to integrate over the relevant coadjoint orbit, whose cotangent directions are parametrized by the ghost zero modes. One can integrate objects taking values on the canonical line bundle (i.e.\ top forms). Therefore, the ghost partition functions must take values in the inverse line bundle.

We use the following fact repeatedly. A fermionic $\mathfrak{b}\mathfrak{c}$ ghost system with $h(\mathfrak{b})=\frac{1}{2}+\lambda$ and $h(\mathfrak{c})=\frac{1}{2}-\lambda$ has central charge $1-12\lambda^2$ and a bosonic $\beta\gamma$ superghost system with $h(\beta)=\frac{1}{2}+\lambda$ and $h(\gamma)=\frac{1}{2}-\lambda$ has central charge $-1+12 \lambda^2$ \cite{Friedan:1985ge}.

\subsection{Virasoro algebra} \label{subapp:virasoro-ghosts}
To warm up, let us recall the well-known case of the Virasoro algebra. The corresponding ghost-system are the string theory $\mathfrak{b}\mathfrak{c}$-ghosts where $\mathfrak{b} \equiv \mathfrak{b}_{zz}$ is a quadratic differential and $\mathfrak{c} \equiv \mathfrak{c}^z$ a vector field. They famously generate a Virasoro algebra of central charge $c=-26$. Therefore the canonical line bundle is identified as $\mathcal{K}_{\mathcal{O}} \cong \mathcal{L}_{c=26}$. 

\subsection{\texorpdfstring{$\mathcal{N}=1$}{N=1} Virasoro algebra} \label{subapp:n1-ghosts}
This case parallels the RNS superstring. One introduces the reparametrization ghosts $(\mathfrak{b},\mathfrak{c})$ with
$h(\mathfrak{b})=2$ together with a bosonic superghost system $(\beta,\gamma)$
with $h(\beta)=\frac{3}{2}$. They generate an $\mathcal{N}=1$ Virasoro algebra of central charge $c=-26+11=-15$. Therefore the Berezinian line bundle satisfies $\mathcal{K}_{\mathcal{O}} \cong \mathcal{L}_{c=15}$.

\subsection{Affine Kac--Moody algebra} \label{subapp:affine-ghosts}
Let $\widehat{\mathfrak{g}}$ be an affine Kac--Moody algebra.
To gauge the currents $J^a$ one introduces fermionic ghosts in the adjoint representation
with weights $h(\mathfrak{b}^a)=1$ and $h(\mathfrak{c}^a)=0$, so that the BRST current
starts as $\mathfrak{c}_a J^a+\dots$, where the omitted terms are trilinear in the ghosts.
The ghost bilinears generate an affine Kac--Moody algebra realized by $j_{\mathfrak{b}\mathfrak{c}}^a=\tensor{f}{^a_{bc}}\mathfrak{b}^b\mathfrak{c}^c$. $\mathfrak{b}^a$ and $\mathfrak{c}^a$ have the same defining OPE as complex fermions in the adjoint representation. Real fermions in the adjoint representation are well-known to generate a current algebra of level equal to the dual Coxeter number, $k=h^\vee$. Therefore, complex fermions generate a current algebra of level $k=2h^\vee$. 
When one also keeps track of the finite-dimensional weight data (for regular orbits with a chosen
polarization), the same determinant yields the familiar $2\rho$-shift in the finite part,
with $\rho$ the Weyl vector. 
Therefore, $\mathcal{K}_{\mathcal{O}} \cong \mathcal{L}_{-2\rho,-2h^\vee}$, which is the result we used in Section~\ref{subsec:affine-kac-moody}. 

\subsection{\texorpdfstring{$\mathcal{N}=2$}{N=2} Virasoro algebra} \label{subapp:n2-ghosts}
For the $\mathcal{N}=2$ super Virasoro algebra (relevant for $\mathcal{N}=2$ strings),
one gauges $T$, the two supercurrents $G^\pm$, and the $\U(1)$ R-current $J$.
Accordingly, one introduces:
\begin{itemize}
\item A ghost system $(\mathfrak{b},\mathfrak{c})$ with $h(\mathfrak{b})=2$, $h(\mathfrak{c})=-1$,
\item a pair of superghost systems $(\beta^\pm,\gamma^\pm)$ with
$h(\beta^\pm)=\frac{3}{2}$, $h(\gamma^\pm)=-\frac{1}{2}$,
\item an additional $(\mathfrak{b}',\mathfrak{c}')$ system for the $\U(1)$ gauge symmetry
with $h(\mathfrak{b}')=1$, $h(\mathfrak{c}')=0$.
\end{itemize}
Constructing the full $\mathcal{N}=2$ ghost superconformal algebra requires one to bosonize the superghosts and working in the `large Hilbert space';
for the purpose of determining $\mathcal{K}_{\mathcal{O}}$ only the anomaly coefficients
are needed and are independent of this choice.
The total ghost central charge is
\be 
c=-26+2 \cdot 11-2=-6\ .
\ee
In the worldsheet interpretation this is the standard $c_{\text{matter}}=6$ condition,
i.e.\ two complex target dimensions (or four real dimensions with split signature)
\cite{Ooguri:1991fp}. Therefore, $\mathcal{K}_{\mathcal{O}} \cong \mathcal{L}_{c=6}$.

\subsection{\texorpdfstring{$\mathcal{N}=4$}{N=4} Virasoro algebra} \label{subapp:n4-ghosts}
For the (small) $\mathcal{N}=4$ super Virasoro algebra one gauges $T$, four supercurrents $G^{\alpha\dot\alpha}$,
and an $\su(2)_{\mR}$ current algebra.
The corresponding ghost content is:
\begin{itemize}
\item $(\mathfrak{b},\mathfrak{c})$ with $h(\mathfrak{b})=2$, $h(\mathfrak{c})=-1$,
\item four bosonic superghosts $(\beta^{\alpha\dot\alpha},\gamma^{\alpha\dot\alpha})$ with
$h(\beta^{\alpha\dot\alpha})=\frac{3}{2}$, $h(\gamma^{\alpha\dot\alpha})=-\frac{1}{2}$,
\item $(\mathfrak{b}^a,\mathfrak{c}^a)$ in the adjoint of $\su(2)_{\mR}$
with $h(\mathfrak{b}^a)=1$, $h(\mathfrak{c}^a)=0$.
\end{itemize}
The total central charge is
\begin{equation}
c
=
-26 + 4\cdot 11 - 3\cdot 2
=
12\ ,
\label{eq:n4-ghost-central-charge}
\end{equation}
and thus the Berezinian line bundle satisfies $\mathcal{K}_{\mathcal{O}} \cong \mathcal{L}_{c=-12}$.
As a useful consistency check, the $\su(2)_{\mR}$ currents built from
ghost bilinears have the expected level.
The adjoint $(\mathfrak{b}^a,\mathfrak{c}^a)$ ghosts generate $\su(2)$ at level
$k_{\mathfrak{b}\mathfrak{c}}=2h^\vee=4$.
Contracting the outer automorphism index $\dot\alpha$, the $(\beta^{\alpha\dot\alpha},\gamma^{\alpha\dot\alpha})$
systems furnish two bosonic doublets of $\su(2)_{\mR}$, hence contribute
$k_{\beta\gamma}=-2$.
Altogether one obtains an $\su(2)_{k}$ subalgebra with $k=2$, and for the small
$\mathcal{N}=4$ algebra one has $c=6k$ \cite{Eguchi:1987sm},
in agreement with \eqref{eq:n4-ghost-central-charge}.

\bibliographystyle{JHEP}
\bibliography{bib}
\end{document}